\documentclass[letterpaper,11pt]{article}
\pdfoutput=1
\usepackage{jheppub}
\usepackage{slashed}
\usepackage{graphicx}
\usepackage{subfigure}
\usepackage{soul}
\usepackage{amsmath}
\usepackage{multirow}
\usepackage{tablefootnote}
\usepackage{hyperref}
\usepackage{xcolor}
\usepackage{booktabs}
\usepackage{epstopdf}
\usepackage{lscape}
\usepackage{url}

\usepackage[dvipsnames]{xcolor}
\usepackage[utf8]{inputenc}

\newcommand{\beq}{\begin{equation}}
\newcommand{\eeq}{\end{equation}}
\newcommand{\bea}{\begin{eqnarray}}
\newcommand{\eea}{\end{eqnarray}}

\def\m1{M_1}
\def\m2{M_2}
\def\m3{M_3}

\def\ch10{\tilde \chi^0_1}

\def\to{\rightarrow}

\newcommand{\lsim}{\mathrel{\mathop{\kern 0pt \rlap
  {\raise.2ex\hbox{$<$}}}
  \lower.9ex\hbox{\kern-.190em $\sim$}}}
\newcommand{\gsim}{\mathrel{\mathop{\kern 0pt \rlap
  {\raise.2ex\hbox{$>$}}}
  \lower.9ex\hbox{\kern-.190em $\sim$}}}

\definecolor{pink}{RGB}{255,105,180}

\definecolor{green2}{rgb}{0,0.56,0.32}

\def\figureautorefname~#1\null{Fig.\,#1\null}
\def\tableautorefname~#1\null{Tab.\,#1\null}

\def\equationautorefname~#1\null{Eq.\,(#1)\null}

\newcommand{\tanb}{\tan \beta}

\title{Probing the 2HDMS through Higgs Precision Measurements}

\author[a]{Cheng Li}

\author[a]{, Wei Su}
\author[a]{ and Huang Zeng\footnote{First author.}}

\affiliation[a]{School of Science, Shenzhen Campus of Sun Yat-sen University, No. 66, Gongchang Road, \\ Guangming District, Shenzhen, Guangdong 518107, China }

\emailAdd{lich389@mail.sysu.edu.cn}

\emailAdd{suwei26@mail.sysu.edu.cn}

\emailAdd{zengh67@mail2.sysu.edu.cn}

\abstract{
Precision measurements of the 125 GeV Higgs boson provide a powerful probe of physics beyond the Standard Model. In this work, we investigate the sensitivity of current and future collider experiments, including the LHC, HL-LHC, CEPC, ILC, and FCC-ee, to the singlet-extended two-Higgs-doublet model (2HDMS). We perform global fits to Higgs precision observables for all four Yukawa types, deriving constraints on the doublet alignment parameter $\cos(\beta-\alpha)$, $\tan\beta$, and the singlet-doublet mixing angle $\alpha_{hS}$ at the 95\% C.L.. Typically $|\cos(\beta-\alpha)|$ at Type-II and Type-I under LHC are less than 0.043 and 0.32, and change little with $\alpha_{hS}$.
We also find that the latest LHC measurements already provide stronger constraints in certain regions than the HL-LHC 2019 projections. We therefore present updated HL-LHC sensitivity estimates extrapolated from current LHC measurements, obtaining approximately a factor-of-two improvement in the constraints on $|\cos(\beta-\alpha)|$ for Types-II, -L, and -F relative to the earlier projections. 
For scenarios with $m_{h_S}< m_h/2$, we find that exotic Higgs decays, $h\to h_S h_S$, can provide stronger constraints than Higgs precision fits and conventional searches in the large-$\tan\beta$ region of Type-I models. The allowed $\lvert \alpha_{hS} \rvert$ is shrunk from 0.035 without light Higgs under HL-LHC to be less than 0.001 here . Our results highlight the complementarity between precision Higgs measurements and exotic decay searches in probing the extended Higgs sector at current and future colliders.
}


\begin{document}
\maketitle
\flushbottom
\section{Introduction}
\label{sec1}

The discovery of a Standard Model (SM)-like Higgs boson with a mass of approximately 125 GeV at the Large Hadron Collider (LHC) in 2012 by the ATLAS and CMS collaborations \cite{ATLAS:2012, CMS:2012} marks a pivotal milestone in particle physics, confirming the Higgs mechanism as the origin of electroweak symmetry breaking. Despite its extraordinary success, the SM is widely recognized as incomplete, leaving unresolved puzzles such as the nature of dark matter, the baryogenesis, neutrino masses, etc. \cite{Bertone:2004, Cohen:1993, Georgi:1974}. These shortcomings motivate extensive searches for physics beyond the SM (BSM), and many compelling extensions introduce an enlarged Higgs sector to address these issues. So far, the LHC has performed a dedicated measurement for the 125 GeV Higgs boson properties, which are presented in the Run-I combined results \cite{ATLAS:2016neq} and Run-II results of $WW$~\cite{ATLAS:2019vrd,CMS:2020dvg}, $ZZ$~\cite{ATLAS:2020rej,CMS:2021ugl}, $bb$~\cite{bbATLAS,bbCMS}, $\gamma\gamma$~\cite{gammaCMS,gammaATLAS}, $\tau\tau$~\cite{tautauATLAS,tautauCMS}, $\mu\mu$~\cite{muATLAS,muCMS}, $cc$~\cite{ccATLAS,ccCMS} and $gg$~\cite{gg} channels. Based on these experimental results, one can determine the exclusion bound of the 125 GeV Higgs measurement at 95\% C.L. in previous study~\cite{ATLAS:2016neq}, while the current LHC bound can still accommodate the sizable BSM contribution on 125 GeV Higgs.  In order to obtain more hints of the BSM effects, future colliders are motivated, which include the High-Luminosity LHC (HL-LHC)~\cite{HL-LHC,HL-LHC2749422}, and Circular Electron Positron Collider (CEPC)~\cite{CEPC1,CEPC2} and so on. These future machines expect the better precision measurement of 125~GeV Higgs and would potentially measure the new channels that cannot be detected in the current LHC. Therefore, we can estimate the detection capability of future colliders and compare it to the current LHC for a BSM model.

Among the simplest BSM scenarios, the Two-Higgs-Doublet Model (2HDM) is the  most thoroughly studied~\cite{Branco:2011}. It augments the SM with a second Higgs doublet, leading to a physical spectrum comprising two CP-even neutral scalars ($h$, $H$), one CP-odd pseudoscalar ($A$), and a pair of charged Higgs bosons ($H^\pm$). 
A further natural extension is the 2HDM with a complex singlet field, denoted as 2HDM+S \cite{Baum:2018, Heinemeyer:2021}, which not only shares the low-energy Higgs sector of the Next-to-Minimal Supersymmetric Standard Model (NMSSM) \cite{Ellwanger:2009} but also offers a viable dark matter candidate \cite{Dutta:2023} and can accommodate the intriguing 95 GeV excess observed at LEP and the LHC \cite{Biekotter:2023}. The 2HDM+S enriches the phenomenology further, introducing three CP-even ($h$, $H$, $h_S$), two CP-odd ($A$, $A_S$), and a pair of charged Higgs bosons ($H^\pm$), thereby opening up a wide variety of decay channels and collider signatures.

As the LHC measurement proceed, several studies have explored the LHC experimental constraints on the 2HDM parameter space \cite{Celis:2013rcs,Craig:2013hca,Eberhardt:2013uba,Ferreira:2013qua,Bernon:2014nxa,Bernon:2015qea,Bernon:2015wef,Chowdhury:2015yja,Bian:2017gxg,Su:2019,Heo:2025vkz}, in particular the $\tan\beta$ vs $\cos(\beta-\alpha)$ parameter space, where these two parameters dominate the properties of 125~GeV Higgs boson. However, in the 2HDMS, the additional singlet admixture enters into 125~GeV Higgs boson, and three of parameters can determine the properties of 125~GeV Higgs boson.
Previous phenomenological studies of the 2HDM+S have often been restricted to specific symmetry-imposed scenarios. In this work, we adopt a more general and model-independent parametrization using physical Higgs masses and mixing angles, allowing our results to be readily mapped onto various concrete realizations with different underlying symmetries. To systematically disentangle the complex interplay of multiple mixing angles, we consider five benchmark cases where at most one mixing angle is non-zero, as defined in our earlier electroweak precision analysis \cite{Li:2025a}. This strategy enables us to isolate the distinct impact of each mixing parameter on experimental observables. Based on this mixing framework, we mainly explore the impact of the corresponding mixing angles, which can determine the experimental properties of 125~GeV Higgs in this paper.

The structure of this paper is as follows. In Section.2, we will briefly introduce some properties of the 2HDMS model to facilitate understanding of this framework. In Section.3, we will describe the methods employed in the study of 2HDMS, including the relevant experimental results. Specially, we have performed an extrapolation of the experimental precision for the HL-LHC and presented the resulting values. We will also provide a brief overview of several future lepton colliders. In Section.4, we present the results of the 2HDMS model obtained using multiple colliders, including those derived from HL-LHC data with extrapolation accuracy. We analyze the reached preliminary conclusions. In Section.5, we analyze the scenario in which, beyond the SM-like Higgs particle, there exists a low mass Higgs particle within the 2HDMS, and discuss its implications for the parameter space of 2HDMS. In Section.6, we employ multiple electron-positron colliders and compare their differences under various luminosities conditions at high energies. Section.7 serves as a summary. we will review the entire work of this article and draw conclusions. 
 
\section{Introduction to 2HDMS}
\label{sec2}

The 2HDMS, the 2HDM with the singlet extension S, has the following scalar fields content
\begin{equation}
        \phi_1 = \left(\begin{array}{cc}
        \chi_1^+ \\ \phi_1
        \end{array} \right)=\left(\begin{array}{cc}
        \chi_1^+ \\ \frac{v_1+ \rho_1+i\eta_1}{\sqrt{2}}
        \end{array} \right)\ \ \ \ 
        \phi_2 = \left(\begin{array}{cc}
        \chi_2^+ \\ \phi_2
        \end{array} \right)=\left(\begin{array}{cc}
        \chi_2^+ \\ \frac{v_2+ \rho_2+i\eta_2}{\sqrt{2}}
        \end{array} \right),
\end{equation}
\begin{align}
    \begin{split}
S=v_\text{S}+ \rho_\text{S}+i\eta_\text{S}.
    \end{split}
\end{align}
where $\phi_1$,$\phi_2$ are the $\text{SU}(2)_\text{L}$×$\text{U}(1)_\text{Y}$ gauge doublets, and S is the gauge singlet field.

After electroweak symmetry breaking, the neutral components of $\Phi_1$, $\Phi_2$ and $S$ develop  non-zero vacuum expectation values, $v_1$, $v_2$ and $v_S$, with $\sqrt{v_1^2 + v_2^2} = v\approx 246$~GeV.  We also introduce $\tan\beta = \frac{v_2}{v_1}$.

The 2HDMS Higgs potential is the combination of 2HDM part and the singlet S part
\begin{align}
    \begin{split}
 V_{\text{2HDMS}}=V_{\text{2HDM}}+V_\text{S},
    \end{split}
\end{align}
which can be generally expressed as follows:
\begin{align}
    \begin{split}
V_{\text{2HDMS}}&=m^2_{11}\phi_1^\dagger\phi_1+m^2_{22}\phi_2^\dagger\phi_2-(m^2_{12}
\phi_1^\dagger\phi_2+h.c.)\\ &+\lambda_1(\phi_1^\dagger\phi_1)^2/2+\lambda_2(\phi_2^\dagger\phi_2)^2/2+\lambda_3(\phi_1^\dagger\phi_1)(\phi_2^\dagger\phi_2)+\lambda_4(\phi_1^\dagger\phi_2)(\phi_2^\dagger\phi_1)\\& +(\phi_1^\dagger\phi_2)[\lambda_5(\phi_1^\dagger\phi_2)/2+\lambda_6(\phi_1^\dagger\phi_1) +\lambda_7(\phi_2^\dagger\phi_2) +h.c.]\\ &+m_S^2S^\dagger S+m_S^{'2}/2(S^2+h.c.)\\&+(\mu_{S1}S^3/6+\mu_{S2}SS^\dagger S/2+h.c.)+(\lambda_1''S^4/24+\lambda_2''S^2S^\dagger S/6+h.c.)+\lambda_3''(S^\dagger S)^2/4\\&+[S(\mu_{11}\phi_1^\dagger \phi_1+\mu_{22}\phi_2^\dagger \phi_2+\mu_{12}\phi_1^\dagger \phi_2+\mu_{21}\phi_2^\dagger\phi_1)+h.c.]\\&+S^\dagger S[\lambda'_1\phi_1^\dagger\phi_1+\lambda_2'\phi_2^\dagger\phi_2+\lambda_3'\phi_1^\dagger\phi_2+h.c.]\\&+[S^2(\lambda_4'\phi_1^\dagger\phi_1+\lambda_5'\phi_2^\dagger\phi_2+\lambda_6'\phi_1^\dagger\phi_2+\lambda_7'\phi_2^\dagger\phi_1)+h.c.].   \end{split}
\end{align}
It can be seen that this potential contains many parameters, which can cause significant complications when studying the Higgs field potential. To address this issue and suppress the flavor changing neutral current (FCNC), we impose the $Z_2$ symmetries $\phi_1$ $\rightarrow$ -$\phi_1$ between the scalar field and then the $\lambda_6$, $\lambda_7$, $\lambda_3'$, $\lambda_6'$, $\lambda_7'$ terms can be eliminated. Eventually, we obtain the Higgs field potential for 2HDMS as follows:
\begin{align}
\begin{split}
V&=m^2_{11}\phi_1^\dagger\phi_1+m^2_{22}\phi_2^\dagger\phi_2+\lambda_1(\phi_1^\dagger\phi_1)^2/2+\lambda_2(\phi_2^\dagger\phi_2)^2/2+\lambda_3(\phi_1^\dagger\phi_1)(\phi_2^\dagger\phi_2)\\&+\lambda_4(\phi_1^\dagger\phi_2)(\phi_2^\dagger\phi_1) + m_S^{2}S^\dagger S+S^\dagger S(\lambda'_1\phi_1^\dagger\phi_1+\lambda_2'\phi_2^\dagger\phi_2)\\&+\lambda_3''(S^\dagger S)^2/4+(-m^2_{12}\phi^\dagger_1\phi_2+\mu_{S1}S^3/6+\mu_{12} S\phi^\dagger_1\phi_2+h.c.).  \end{split}
\end{align}
There are 12 free parameters, which are tan$\beta$, $\lambda_1$, $\lambda_2$, $\lambda_3$, $\lambda_4$, $\lambda_1'$, $\lambda_2'$, $\lambda_3''$, $m^2_{12}$, $\mu_{S1}$, $\mu_{12}$ and $v_S$. 

Taking the second derivative of the scalar field potential of 2HDMS, we obtain the Higgs mass matrix for the CP-even Higgs bosons as follows:
\begin{align}
    \begin{split}
    M^2_{S11}&=2\lambda_1v^2\cos^2\beta+(m^2_{12}-\mu^2_{12}v_S)\tan\beta,\\ 
     M^2_{S22}&=2\lambda_2v^2\sin^2\beta+(m^2_{12}-\mu^2_{12}v_S)\cot\beta,\\
      M^2_{S12}&=(\lambda_3+\lambda_4)v^2\sin^2\beta-(m^2_{12}-\mu^2_{12}v_S),\\
       M^2_{S13}&=(2\lambda_1'v_S\cos\beta+\mu^2_{12}\sin\beta)v,\\
       M^2_{S23}&=(2\lambda_2'v_S\sin\beta+\mu^2_{12}\cos\beta)v,\\
       M^2_{S33}&=\frac{\mu_{S1}}{2}v_S+\lambda_3''v^2_S-\mu_{12}\frac{v^2}{2v_S}\sin(2\beta).
   \end{split}
\end{align}
To obtain the mass eigenstates of the CP-even Higgs bosons, we diagonalize their mass matrix $M^2_S$. Since the mass matrix itself is symmetric and the diagonalizing matrix must be orthogonal, we use a 3×3 rotation matrix to achieve the diagonalization of the mass matrix. For the CP-even sector, the rotation matrix used is as follows:
\begin{equation}
R = \left(\begin{array}{ccc}
c_{\alpha}c_{\alpha_{HS}}&s_{\alpha}c_{\alpha_{HS}}&s_{\alpha_{HS}} \\ -s_{\alpha}c_{\alpha_{hS}}-c_{\alpha}s_{\alpha_{HS}}s_{\alpha_{hS}} &c_{\alpha}c_{\alpha_{hS}}-s_{\alpha}s_{\alpha_{HS}}s_{\alpha_{hS}}&c_{\alpha_{HS}}s_{\alpha_{hS}}\\
s_{\alpha}s_{\alpha_{hS}}-c_{\alpha}s_{\alpha_{HS}}c_{\alpha_{hS}}
&-s_{\alpha}s_{\alpha_{HS}}c_{\alpha_{hS}}-c_{\alpha}s_{\alpha_{hS}}&c_{\alpha_{HS}}c_{\alpha_{hS}}      \end{array} \right) ,
    \end{equation}
with $s_x = \sin x, c_x =\cos x$, and 
\begin{equation}
        R = \left(\begin{array}{ccc}
         \rho_1\\  \rho_2\\
         \rho_3     \end{array} \right) =\left(\begin{array}{ccc}
         H\\  h\\
         h_S   \end{array} \right),
\end{equation}
\begin{align}
    \begin{split}
    M^2_S=R^Tdiag \{m^2_{H},m^2_{h},m^2_{h_S}\}R.
   \end{split}
\end{align}
Here, the three mixing angles are the angle $\alpha$ for the mixing between two 2HDM CP-even fields, the angle $\alpha_{hS}$ for the mixing between the SM-like Higgs and the singlet field, and the angle $\alpha_{HS}$ for the mixing between the addtional 2HDM Higgs and singlet field. Therefore, the three CP-even Higgs bosons are the additional doublet-like Higgs boson H, the SM-like Higgs boson h and the singlet-like Higgs boson $h_S$. 

The Lagrangian for the Yukawa coupling terms in the 2HDM is expressed as :
\begin{align}
    \begin{split}
        -\mathcal{L}_Y =Y_\mu\overline{\mu}_L\mu_R\phi_i+Y_d\overline{d}_Ld_R\phi_j+Y_\iota\overline{\iota}_L\iota_R\phi_k+h.c..
    \end{split}
\end{align}
where different combinations of $\phi_i$ , $\phi_j$ , $\phi_k$ represent different types of Yukawa couplings.

The couplings of the neutral scalar Higgs to the particle is denoted as $g_{h_iff}$. We define the reduced coupling as the ratio between the 2HDMS Higgs coupling and the corresponding SM-Higgs coupling:
\begin{align}
    \begin{split}
    c_{h_iff}=\frac{g_{h_iff}}{g_{h_\text{SM}ff}}.
    \end{split}
\end{align}

As the 2HDM, the Higgs to fermions couplings of 2HDMS are given by four types of Yukawa couplings which shown as \autoref{tab:4type}. 
\begin{table}[h!] 
    \centering
    \begin{tabular}{*{5}{c|cccc}}
        \toprule
         & Type-I &Type-II& Type-L& Type-F                     \\    
        \midrule
        $c_{{h_i}uu}$ &$R_{i2}/sin\beta $&$R_{i2}/sin\beta$&$R_{i2}/sin\beta $&$R_{i2}/sin\beta $\\
         $c_{{h_i}dd}$&$R_{i2}/sin\beta $&$R_{i1}/cos\beta$&$R_{i2}/sin\beta $&$R_{i1}/cos\beta$\\
         
          $c_{{h_i}ll}$&$R_{i2}/sin\beta $&$R_{i1}/cos\beta$&$R_{i1}/cos\beta $&$R_{i2}/sin\beta$\\
          \midrule
          
    \end{tabular}
     \caption{Reduced Higgs-fermion couplings for four different types of Yukawa couplings. 
    }
    \label{tab:4type}
\end{table}

For the SM-Like Higgs in 2HDMS, we provide the explicit expressions, as \autoref{eq:chvv} and also present the coupling relationships under Type-II as \autoref{eq:chuu}. Regarding the coupling between the SM-Like Higgs boson and other particles in 2HDMS under Type-II, $c_{hVV}$, $c_{huu}$, $c_{hdd}$ and $c_{hll}$ will equals 1 if $\cos(\beta-\alpha)=0$, $\alpha_{HS}=\alpha_{hS}=0$. In this case, 2HDMS will degrade to SM. To be clear, here we have 
\begin{align}
    \begin{split}
        c_{hVV}=\sin(\beta-\alpha)\cos(\alpha_{hS}) - \cos(\beta-\alpha)\sin(\alpha_{HS})\sin(\alpha_{hS}),  \end{split}
        \label{eq:chvv}
\end{align}
\begin{align}
    \begin{split}
        c_{huu}&=c_{hVV}+[\cos(\beta-\alpha)\cos(\alpha_{hS})+\sin(\beta-\alpha)\sin(\alpha_{hS})\sin(\alpha_{HS})]/\tanb, \\
        c_{hdd}&=c_{hVV}-[\cos(\beta-\alpha)\cos(\alpha_{hS})+\sin(\beta-\alpha)\sin(\alpha_{hS})\sin(\alpha_{HS})]\tanb, \\
         c_{hll}&=c_{hVV}-[\cos(\beta-\alpha)\cos(\alpha_{hS})+\sin(\beta-\alpha)\sin(\alpha_{hS})\sin(\alpha_{HS})]\tanb.
         \label{eq:chuu}
        \end{split}
\end{align}
Under the subsequent constraints of the 2HDMS parameter space in Type-II, they can be further simplified to some benchmark cases.
\begin{itemize}
\item $\tan\beta$ vs $\cos(\beta-\alpha)$ with $\alpha_{hS}=\alpha_{HS}=0$
\begin{align}
    \begin{split}
        c_{hVV}&=\sin(\beta-\alpha) ,\\
        c_{huu}&=c_{hVV}+\cos(\beta-\alpha)/\tanb, \\
        c_{hdd}&=c_{hVV}-\cos(\beta-\alpha)\tanb, \\
         c_{hll}&=c_{hVV}-\cos(\beta-\alpha)\tanb.
         \label{eq:coupling1}
        \end{split}
\end{align}
It will mainly discuss the mixing effects between the two CP-even Higgses.
\item $\alpha_{hS}$ vs $\cos(\beta-\alpha)$ with $\alpha_{HS}=0$ and $\tanb=1$
\begin{align}
    \begin{split}
        c_{hVV}&=\sin(\beta-\alpha)\cos(\alpha_{hS}),\\
        c_{huu}&=c_{hVV}+\cos(\beta-\alpha)\cos(\alpha_{hS}), \\
        c_{hdd}&=c_{hVV}-\cos(\beta-\alpha)\cos(\alpha_{hS}), \\
         c_{hll}&=c_{hVV}-\cos(\beta-\alpha)\cos(\alpha_{hS}).
         \label{eq:coupling2}
        \end{split}
\end{align}
\item $\alpha_{hS}$ vs $\tanb$ with $\alpha_{HS}=0$ and $\cos(\beta-\alpha)=0$
\begin{align}
    \begin{split}
        c_{hVV}&=\cos(\alpha_{hS}),\\
        c_{huu}&=c_{hdd}=c_{hll}=c_{hVV}. \\
         \label{eq:coupling3}
        \end{split}
\end{align}
\end{itemize}
They are for the mixing effects between the CP-even Higgs and singlet one.
%
\section{Collider phenomenology}
\label{sec3}
%
\subsection{Global fit framework}

We use the HiggsTools\cite{higgstool} to study and analyze the 2HDMS model. 

\begin{figure}[h]
    \centering
    \includegraphics[width=0.5\linewidth]{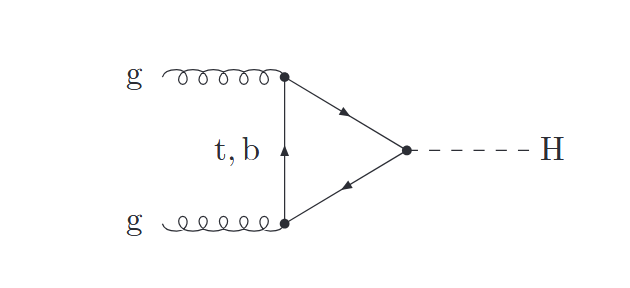}
    \caption{Feynman diagram of ggF at lowest order\cite{LHCHiggsCrossSectionWorkingGroup:2011wcg}.}
    \label{fig:ggf}
\end{figure}

\begin{figure}[h]
    \centering
    \includegraphics[width=0.8\linewidth]{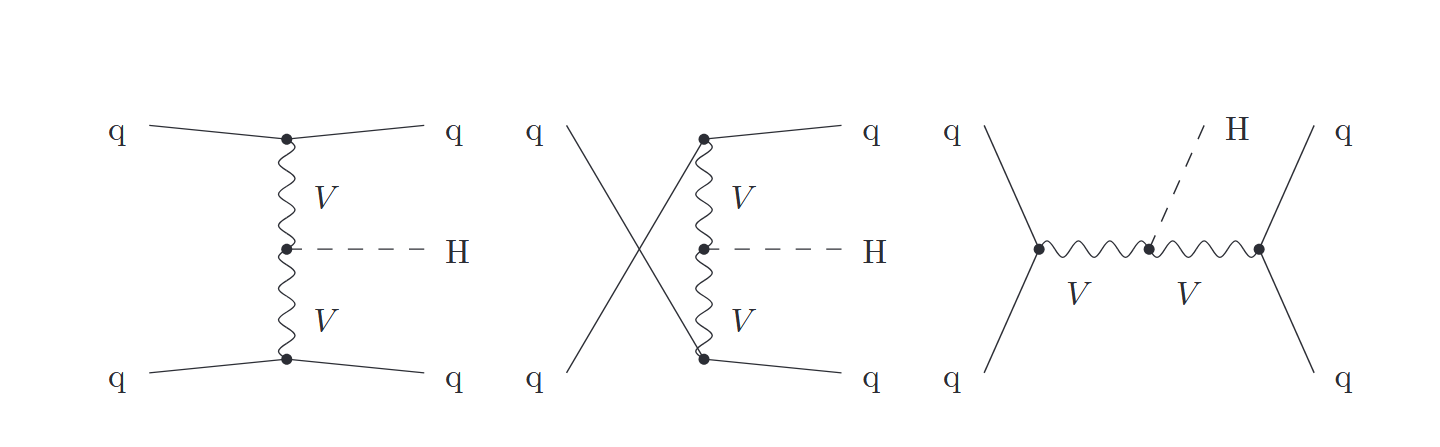}
    \caption{Feynman diagram of VBF\cite{LHCHiggsCrossSectionWorkingGroup:2011wcg}.}
    \label{fig:vbf}
\end{figure}

\begin{figure}[h]
    \centering
    \includegraphics[width=0.8\linewidth]{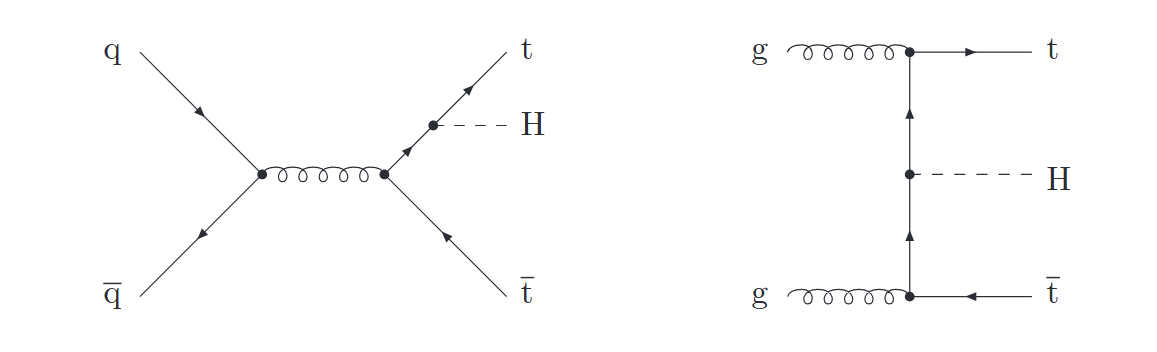}
    \caption{Feynman diagram of ttH\cite{LHCHiggsCrossSectionWorkingGroup:2011wcg}.}
    \label{fig:tth}
\end{figure}

We employ three colliders (LHC, HL-LHC and CEPC) to investigate the parameter space of 2HDMS. The LHC is a proton-proton collider that achieved tremendous success during its Run 1, during which the Higgs boson was discovered in 2012. Run 2 focused on precision measurements, increasing the energy of proton-proton collisions from 7–8 TeV (Run 1) to 13 TeV (Run 2). Throughout Run 2, both the ATLAS and CMS experiments collected approximately luminosity 140 $fb^{-1}$ . The data we utilize comes exclusively from LHC Run 2. For the LHC at 13 TeV, the production channels for Higgs bosons are as follows: ggF, VH (V = W, Z), VBF and ttH where Higgs decay channels include $\gamma\gamma$, $ZZ$, $WW$, $\tau\tau$, $bb$, $cc$, $\mu\mu$ and $gg$. Gluon fusion through a heavy-quark loop is the main production mechanism of the Standard Model Higgs boson at hadron colliders, and we display the lowest-order Feynman diagram of ggF, as \autoref{fig:ggf}. Searches for the Higgs boson in the WH and ZH production modes, usually defined as Higgs-strahlung  processes. Here, we display some Feynman diagram of VH, shown as \autoref{fig:wh,zh}. The “vector-boson fusion” (VBF) channel is the production of a Standard Model Higgs boson in association with two hard jets in the forward and backward regions of the detector. \autoref{fig:vbf} shows the Feynman diagram of VBF. The Feynman diagram for ttH is also shown in the \autoref{fig:tth}.
\begin{figure}[h]
    \centering
    \includegraphics[width=0.8\linewidth]{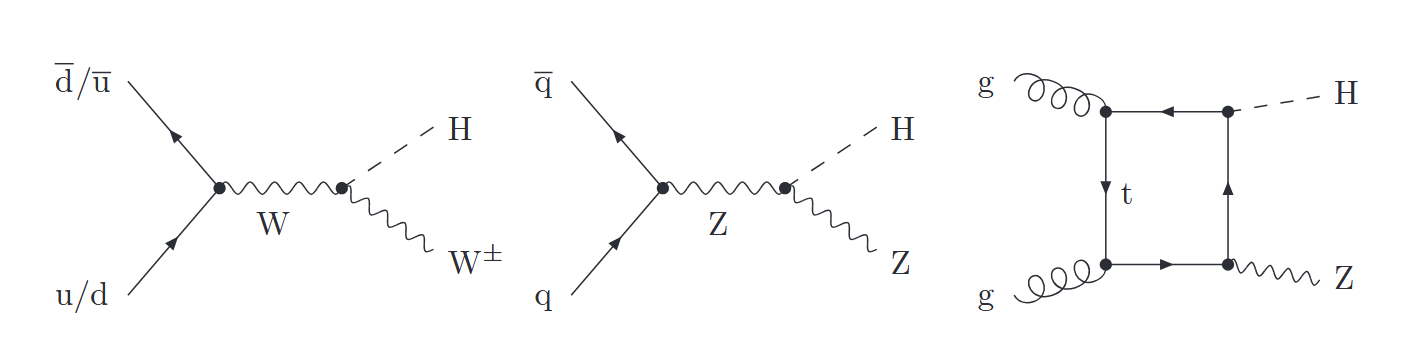}
    \caption{Feynman diagram of VH\cite{LHCHiggsCrossSectionWorkingGroup:2011wcg}.}
    \label{fig:wh,zh}
\end{figure}

For the HL-LHC, at 14 TeV, the production channels for Higgs bosons include ggF, VH, VBF and ttH where Higgs decay channels include $\gamma\gamma$, $ZZ$, $WW$, $\tau\tau$ and $bb$.  As a lepton collider, the CEPC enhances the precision of Higgs boson measurements, aiming to study the properties of the Higgs boson with high accuracy. For CEPC around 240 GeV, the production channel for Higgs bosons is only ZH, and the decay channels include $\gamma\gamma$, $ZZ$, $WW$, $\tau\tau$, $bb$, $cc$, $\mu\mu$ and $gg$.

To convert the experimental precision measurements into constraints on model parameters, we employ global fitting by constructing the $\chi^2$,
\begin{align}
    \begin{split} 
    \chi^2 = \sum_{i}\frac{(\mu^\text{{BSM}}_i-\mu^\text{{obs}}_i)^2}{\sigma^2_{\mu_i}},
     \end{split}
\end{align}
with
\begin{align}
    \begin{split}
     \mu^\text{BSM}_i = \frac{(\sigma \times Br)_\text{{BSM}}}{(\sigma \times Br)_\text{{SM}}} = c^2_{h_iff} \times \frac{Br_\text{BSM}}{Br_\text{SM}}.
    \end{split}
\end{align}

Here $\mu^\text{{BSM}}_i$ is determined by each Higgs boson production channel and its corresponding decay channel, defined as the ratio of the production cross section multiplied by the corresponding decay branch ratio of 2HDMS to them of the Standard Model. $\sigma_{\mu_i}$ is the corresponding experimental precision. We note that apart from LHC calculations, the correlations among different channels are usually not provided for multiple colliders including HL-LHC and CEPC, which are thus assumed to be zero.  Further, we make a global fit $\chi^2$ with the profile likelihood method. To ensure the reliability of the results, we take the confidence level (C.L.) for $\chi^2$ at 95\%, which means $\triangle \chi^2 = \chi^2 - \chi^2_{\text{min}} = 5.99 $ for two parameter fit.

We find that, current LHC may work better in some regions than HL-LHC data at 2019\cite{HLLHC:2019}, which may come from better performance at updated LHC, thus we also have the extrapolated study for HL-LHC based on the latest LHC performance. Here, we extrapolate the experimental precision of HL-LHC based on the integrated luminosity values used at both LHC and HL-LHC. The integrated luminosity at LHC is approximately 140 $fb^{-1}$, whereas that at HL-LHC is 3000 $fb^{-1}$.
Thus we can get the extrapolated HL-LHC precision at $\mathcal{L} =3000~fb^{-1}$ through
\begin{align}
    \begin{split}
    \Delta\sigma^\text{{HL-LHC}} = \Delta\sigma^\text{{LHC}} \times \sqrt{\frac{\mathcal{L}^\text{{LHC}}}{\mathcal{L}^\text{{HL-LHC}}}}.
 \end{split}
\end{align}

Finally we can utilize the extrapolated HL-LHC precision in our research. As shown in \autoref{fig:cbatb} in Section.4, HL-LHC results with extrapolation accuracy will yield significantly better outcomes than the original results, while even approaching those obtained by CEPC. This indicates that HL-LHC with extrapolation accuracy will better facilitate our study of the parameter space of 2HDMS. In later sections, we will also continue to present HL-LHC results regarding extrapolation accuracy. 

The various precisions used in the calculation for LHC and HL-LHC are shown in the \autoref{tab:CMS precision} and \autoref{tab:ATLAS precision}. For HL-LHC precision and CEPC precision\cite{CEPC2025wew}, some decay channels only provide the systematic and statistical uncertainties. We therefore combine them in quadrature to obtain the final uncertainty.

\begin{table}[h!]
    \centering
    \begin{tabular}{ccccc}
    \hline
         Higgs decay& production &  LHC  &HL-LHC  &HL-LHC \\
         &&&extrapolation&\\
         \hline
        & ggH &$1.06^{+0.12}_{-0.11}$&$1.00^{+0.026}_{-0.024}$& $1.00^{+0.036}_{-0.036}$  \\
          &VBF&$1.05^{+0.34}_{-0.34}$ &$1.00^{+0.073}_{-0.073}$&$1.00^{+0.081}_{-0.081}$ \\
         $\gamma\gamma$ & WH&$1.45^{+0.54}_{-0.49}$ &$1.00^{+0.116}_{-0.105}$&$1.00^{+0.095}_{-0.095}$\\
          & ZH&$1.20^{+0.69}_{-0.61}$ &$1.00^{+0.148}_{-0.131}$ &$1.00^{+0.139}_{-0.139}$\\
         &ttH&$1.35^{+0.34}_{-0.29}$ &$1.00^{+0.073}_{-0.062}$&$1.00^{+0.072}_{-0.072}$\\
         \hline
         &ggH & $1.03^{+0.13}_{-0.12}$&$1.00^{+0.028}_{-0.026}$& $1.00^{+0.04}_{-0.04}$ \\
         &VBF&$0.38^{+0.52}_{-0.38}$&$1.00^{+0.112}_{-0.082}$&$1.00^{+0.12}_{-0.12}$\\
        $ZZ$&WH&$0.00^{+2.64}_{-0.00}$&$1.00^{+0.566}_{-0.000}$&VH $1.00^{+0.19}_{-0.18}$\\
        &ZH&$7.31^{+4.66}_{-6.12}$&$1.00^{+0.999}_{-1.313}$ & \\
        &ttH&$0.00^{+0.89}_{-0.00}$&$1.00^{+0.191}_{-0.000}$&$1.00^{+0.23}_{-0.20}$\\
         \hline
        &ggH&$0.92^{+0.11}_{-0.11}$&$1.00^{+0.024}_{-0.024}$& $1.00^{+0.05}_{-0.04}$\\
        &VBF&$0.82^{+0.24}_{-0.23}$&$1.00^{+0.051}_{-0.049}$&$1.00^{+0.10}_{-0.09}$\\
          $WW$&WH&$2.38^{+0.73}_{-0.69}$& & \\
          &ZH&$1.88^{+0.74}_{-0.65}$& & \\
          &ttH&$1.45^{+0.38}_{-0.33}$& & \\
         \hline
        &ggH&$0.61^{+0.21}_{-0.20}$&$1.00^{+0.045}_{-0.043}$&$1.00^{+0.12}_{-0.11}$ \\
         
        &VBF&$0.86^{+0.17}_{-0.16}$ &$1.00^{+0.036}_{-0.034}$&$1.00^{+0.07}_{-0.06}$\\
        $\tau\tau$&WH&$1.40^{+0.59}_{-0.57}$&$1.00^{+0.127}_{-0.122}$&VH $1.00^{+0.14}_{-0.14}$\\
        &   ZH&$1.88^{+0.66}_{-0.57}$&$1.00^{+0.142}_{-0.122}$ & \\
        &ttH&$0.33^{+0.42}_{-0.39}$&$1.00^{+0.090}_{-0.084}$&$1.00^{+0.25}_{-0.23}$\\
         \hline
        &ggH&$2.22^{+1.53}_{-1.39}$& & \\
        &VBF&$1.65^{+0.76}_{-0.64}$& & \\
        
        $bb$&WH&$1.28^{+0.37}_{-0.36}$ &$1.00^{+0.079}_{-0.077}$ &$1.00^{+0.068}_{-0.065}$\\
        &ZH&$1.11^{+0.27}_{-0.26}$&$1.00^{+0.058}_{-0.056}$&$1.00^{+0.074}_{-0.069}$\\
        &ttH&$0.39^{+0.27}_{-0.26}$& & \\
         \hline
    \end{tabular}
    \caption{LHC precisions for Higgs boson measurements from CMS Collaboration Group~\cite{ATLAS:2026pdi} and (extrapolated) HL-LHC precisions. For HL-LHC, the precision of the $ZZ$ and $WW$~\cite{HLLHC:2019}, the precision of $\gamma\gamma$~\cite{new_gammagamma}, the precision of $\tau\tau$~\cite{new_tautau} and the precision of $bb$~\cite{new_bb}.}
    \label{tab:CMS precision}
\end{table}
\begin{table}[h!]
    \centering
    \begin{tabular}{ccccc}
    \hline
         Higgs decay& production &  LHC  &HL-LHC   &HL-LHC \\
          &&&extrapolation&\\
         \hline
        & ggH &$1.03^{+0.10}_{-0.10}$ &$1.00^{+0.022}_{-0.022}$ &  $1.00^{+0.036}_{-0.036}$ \\
         &VBF&$1.27^{+0.28}_{-0.25}$&$1.00^{+0.060}_{-0.054}$&$1.00^{+0.081}_{-0.081}$ \\
         $\gamma\gamma$&WH&$1.49^{+0.54}_{-0.49}$&$1.00^{+0.117}_{-0.106}$&  $1.00^{+0.095}_{-0.095}$\\

          &ZH&$-0.22^{+0.61}_{-0.54}$ &$1.00^{+0.132}_{-0.117}
         $& $1.00^{+0.139}_{-0.139}$\\
         
          &ttH&$0.89^{+0.32}_{-0.30}$ &$1.00^{+0.069}_{-0.065}$&$1.00^{+0.072}_{-0.072}$\\
         \hline
         
          &ggH &$0.93^{+0.11}_{-0.10}$ &$1.00^{+0.024}_{-0.022}$&$1.00^{+0.04}_{-0.04}$  \\
         $ZZ$ &VBF&$1.28^{+0.50}_{-0.42}$ &$1.00^{+0.108}_{-0.091}$&$1.00^{+0.12}_{-0.12}$\\
          &VH&$1.47^{+1.14}_{-0.92}$&$1.00^{+0.246}_{-0.199}$ &$1.00^{+0.19}_{-0.18}$\\
          &ttH&$1.87^{+1.71}_{-1.17}$&$1.00^{+0.369}_{-0.253}$&$1.00^{+0.23}_{-0.20}$\\
         \hline
        &ggH&$1.19^{+0.11}_{-0.11}$&$1.00^{+0.024}_{-0.024}$&$1.00^{+0.05}_{-0.04}$\\
          &VBF&$0.96^{+0.21}_{-0.19}$&$1.00^{+0.045}_{-0.041}$&$1.00^{+0.10}_{-0.09}$\\
          $WW$&WH&$0.48^{+0.31}_{-0.30}$& & \\
          &ZH&$1.71^{+0.52}_{-0.45}$& & \\
          &ttH&$0.45^{+0.30}_{-0.30}$& & \\
         \hline
        &ggH&$0.80^{+0.30}_{-0.25}$&$1.00^{+0.065}_{-0.054}$&$1.00^{+0.12}_{-0.11}$\\
        &VBF&$1.04^{+0.18}_{-0.16}$&$1.00^{+0.039}_{-0.035}$&$1.00^{+0.07}_{-0.06}$\\
        $\tau\tau$&WH&$1.49^{+0.47}_{-0.44}$&$1.00^{+0.102}_{-0.095}$&VH $1.00^{+0.14}_{-0.14}$\\
        &ZH&$1.08^{+0.47}_{-0.41}$&$1.00^{+0.102}_{-0.089}$ & \\
        &ttH&$0.30^{+0.43}_{-0.42}$&$1.00^{+0.093}_{-0.091}$&$1.00^{+0.25}_{-0.23}$\\
         \hline
        &VBF+ggH&$0.95^{+0.36}_{-0.35}$& & \\
         
        $bb$&WH&$0.98^{+0.21}_{-0.20}$ &$1.00^{+0.045}_{-0.043}$ &$1.00^{+0.068}_{-0.065}$\\
        
        &$ZH$& $0.88^{+0.22}_{-0.20}$&$1.00^{+0.048}_{-0.043}$ &$1.00^{+0.074}_{-0.069}$ \\
        &ttH&$0.82^{+0.24}_{-0.22}$& & \\
         \hline
    \end{tabular}
    \caption{LHC precisions for Higgs boson measurements from ATLAS Collaboration Group~\cite{CMS:2026zxu} and (extrapolated)HL-LHC precisions. For HL-LHC, the precision of the $ZZ$ and $WW$~\cite{HLLHC:2019}, the precision of $\gamma\gamma$~\cite{new_gammagamma}, the precision of $\tau\tau$~\cite{new_tautau} and the precision of $bb$~\cite{new_bb}.}
    \label{tab:ATLAS precision}
\end{table}

\subsection{$e^+ e^-$ collider }
The Future Circular Collider electron-positron (FCC-ee) is a circular collider located at the European Organization for Nuclear Research (CERN). The International Linear Collider (ILC) is a linear accelerator, with plans to be built in Japan. CEPC, FCC-ee, and ILC are all next-generation $e^+ e^-$ colliders which designed to study particles such as the Higgs boson and top quarks with unprecedented precision. For these colliders, the dominant channel to measure the Higgs boson properties is the Higgsstrahlung process, which is $e^+ e^-$ $\rightarrow$ hZ, where the center of mass energies is around 240$-$250 GeV. Due to the nature of lepton colliders, both the inclusive cross section, $\sigma$(hZ), and the exclusive ones of different Higgs decays in terms of $\sigma$(hZ) $\times$ BR, can be measured to remarkable precisions.

Here, we focus on the differences in results obtained by CEPC, FCC-ee and ILC when the center of mass energy is approximately 240$-$250 GeV. For CEPC, it collects 20 $ab^{-1}$ data at 240 GeV where for FCC-ee is 10.8 $ab^{-1}$ at 240 GeV and for ILC is only 2 $ab^{-1}$ at 250 GeV. For FCC-ee and ILC, the various precisions used~\cite{FCCandILC}.

\section{Study Results} 
\label{sec4}
To explore the contribution of the singlet scalar, we start from heavy $m_{h_S}>m_h/2$, to avoid the additional new exotic decay and find out the mixing effects alone. We note that all the tree-level deviations to the SM couplings in~\autoref{eq:chvv} and~\autoref{eq:chuu} are parametrized by three parameters: $\cos(\beta-\alpha)$, $\tan\beta$ and $\alpha_{hS}$. Therefore, we studied the constraints on the 2HDMS tree level effects in three kinds of parameters space which include $\tan\beta$ vs $\cos(\beta-\alpha)$, $\cos(\beta-\alpha)$ vs $\alpha_{hS}$ and $\tan\beta$ vs $\alpha_{hS}$ with the Higgs precision measurements. Using HiggsTools, we perform a global fit to the Higgs precision measurement at LHC, HL-LHC and CEPC for 4 types of 2HDMS. Here we first consider BSM Higgs is heavier than SM-Like Higgs.

\subsection{$\tan\beta$ - $\cos(\beta-\alpha)$}

\begin{figure}[htpb]
  \centering
\includegraphics[width=0.49 \linewidth]{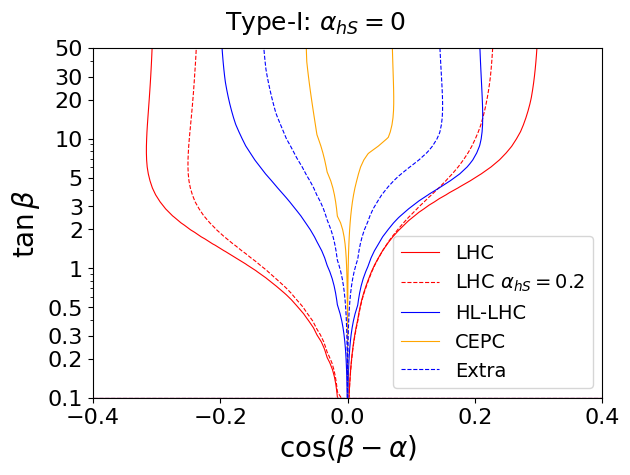}
\includegraphics[width=0.49 \linewidth]{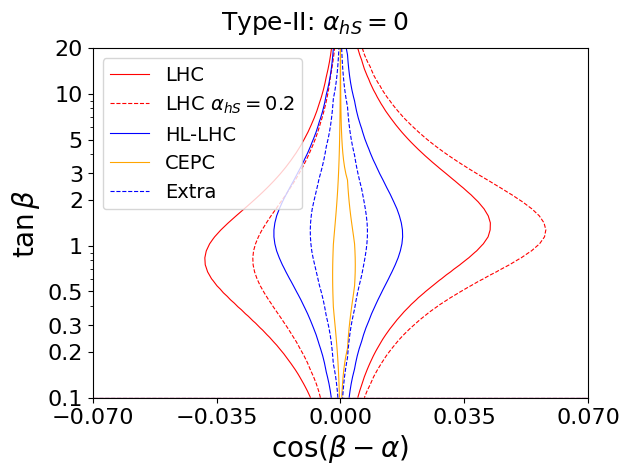}
\includegraphics[width=0.49 \linewidth]{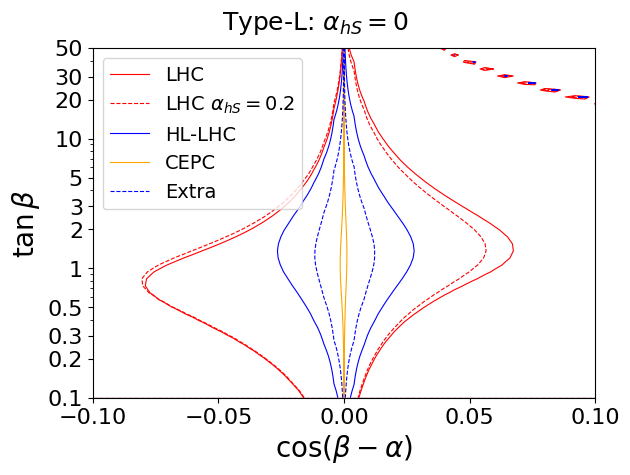}
\includegraphics[width=0.49 \linewidth]{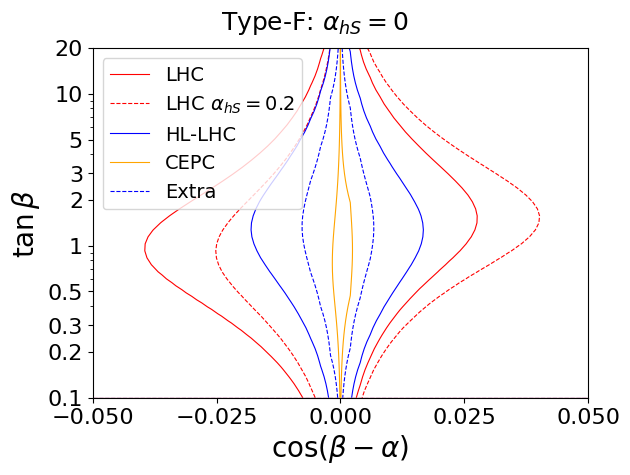}
\caption{The allowed regions at 95\% C.L. for the four types of 2HDMS in the plane of $\tan\beta$ vs $\cos(\beta-\alpha)$. The red solid line is for the current LHC precision measurement at 13 TeV with around 140 $fb^{-1}$ . The blue solid line is for the HL-LHC precision measurement at 14 TeV with 3000 $fb^{-1}$ while the blue dashed line is for the extrapolated HL-LHC. The yellow solid line is for the CEPC precision measurement at a center-of-mass energy of 240 GeV with 20 $ab^{-1}$. All $\alpha_{hS}=0$ here except for the red dashed line for LHC with $\alpha_{hS}=0.2$. }
\label{fig:cbatb}
\end{figure}

We first have \autoref{fig:cbatb} displaying the 95\%C.L. allowed regions in $\tanb$ vs $\cos(\beta-\alpha)$ for various types of 2HDMS parameter space with $\alpha_{hS}=0$. Based on \cite{Li:2605} with $m_{h_S}>m_h/2$, we also set $\alpha_{hS}=0.2$ to analyze the effect of increasing $\alpha_{hS}$ on 2HDMS parameter space and display the result in \autoref{fig:cbatb}. In this case, the SM-like Higgs in 2HDMS is similar to it in 2HDM. 


For the Type-II results of 2HDMS, shown as the upper right panel of \autoref{fig:cbatb}. Noted that for the LHC, HL-LHC, and CEPC, their results in Type-II are all approximately symmetric around $\cos(\beta-\alpha)$ = 0. Since the correction to the down quark and lepton Yukawa couplings receive $\tanb$ enhancement at large value, the allowed parameter space at large $\tanb$ is reduced significantly. For HL-LHC, at large $\tanb$, $\tau\tau$ coupling provides the strongest constraint while for CEPC it is $bb$ coupling. At small $\tanb$, the deviation of the Yukawa couplings from the experimental centre value is proportional to 1/$\tanb$ as shown by~\autoref{eq:coupling2}, resulting a reduced allowed parameter space at small $\tanb$. For HL-LHC, at small $\tanb$, $\gamma\gamma$ coupling provides the strongest constraint while for CEPC it is $gg$ coupling. We have the individual channel constraints at the \autoref{fig:kappahl} and \autoref{fig:kappacepc} at the Appendix to get the strongest channel for each part of the parameter space. Due to the reduced allowed parameter space at large $\tanb$ and small $\tanb$, the interval gets is maximal region around $\tanb=1$, which for LHC (-0.0386, 0.0428), HL-LHC (-0.0192, 0.0177), extrapolated HL-LHC (-0.0087, 0.0078) and CEPC (-0.0024, 0.0043). We have the summarized data at \autoref{tab:cbatb}. With $\alpha_{hS}=0.2$, it is clear that the allowed parameter space of LHC precision measurement will shift to the positive direction of $\cos(\beta-\alpha)$. Similar behavior happen in Type-F.

For the Type-I results of 2HDMS, we have it at the upper left panel of \autoref{fig:cbatb}. At the LHC with 140 $fb^{-1}$, $\lvert \cos(\beta-\alpha) \rvert$ is constrained to be less than about 0.4. At the HL-LHC, the range can be further constrained to be less than 0.2. At the CEPC with 20 $ab^{-1}$, $\lvert \cos(\beta-\alpha) \rvert$ is constrained to be less than about 0.07. At small $\tanb$, it is similar to 2HDMS in Type-II that the allowed parameter space is reduced significantly. For HL-LHC, at small $\tanb$, $\tau\tau$ coupling provides the strongest constraint while for CEPC it is $bb$ coupling and $gg$ coupling. Due to the reduced allowed parameter space at small $\tanb$, the interval gets is maximal region around $\tanb>10$, which for LHC (-0.318, 0.291), HL-LHC (-0.200, 0.207), extrapolated HL-LHC (-0.135, 0.144) and CEPC (-0.069, 0.068). When $\alpha_{hS}=0.2$, different from Type-II, the allowed parameter space of LHC precision measurement will show a slight reduction without shift. Type-L shows similar feature.

The lower panels of \autoref{fig:cbatb} shown the Type-L and Type-F result of 2HDMS. It is similar to Type-II, both the large and small $\tanb$ regions are strongly constrained.  Again the allowed parameter space are approximately symmetric around $\cos(\beta-\alpha)$ = 0 and the interval is maximal around $\tanb=1$.

To better clarify the constraints on the 2HDMS parameter space under this scenario, a summary of the maximum and minimum permissible values of $\cos(\beta-\alpha)$ is given in \autoref{tab:cbatb}. Based on the solid red and blue lines of ~\autoref{fig:cbatb}, we also want to point out that, the upgraded analysis method for LHC is tremendous, making its probe ability reaching to the HL-LHC at 2019 in small and/or large $\tanb$ regions. And the dashed blue line for extrapolated HL-LHC based on latest LHC works much better than HL-LHC 2019 as the~\autoref{tab:cbatb}, which also confirms it.

\begin{table}[h!] 
    \centering
    \begin{tabular}{*{2}{c|cccc}}
        \toprule 
         Type &LHC&HL-LHC&Extra&CEPC   \\   
        \midrule
        Type-I &(-0.318, 0.291)&(-0.200, 0.207)&(-0.135, 0.144)&(-0.069, 0.068)\\
          Type-II &(-0.0386, 0.0428)&(-0.0192, 0.0177)&(-0.0087, 0.0078)&(-0.0024, 0.0043)\\
          Type-L &(-0.079, 0.067)&(-0.027, 0.028)&(-0.012, 0.012)&(-0.002, 0.001)\\
           Type-F &(-0.0397, 0.0277)&(-0.0182, 0.0167)&(-0.0077, 0.0067)&(-0.0018, 0.0024)\\
          \midrule
    \end{tabular}
    
     \caption{Maximum allowed range for $\cos(\beta-\alpha)$ at 95\% C.L. under various colliders.
    }
    \label{tab:cbatb}
\end{table}

\subsection{$\alpha_{hS}$ - $\cos(\beta-\alpha)$}
\begin{figure}[htpb]
  \centering
\includegraphics[width=0.49 \linewidth]{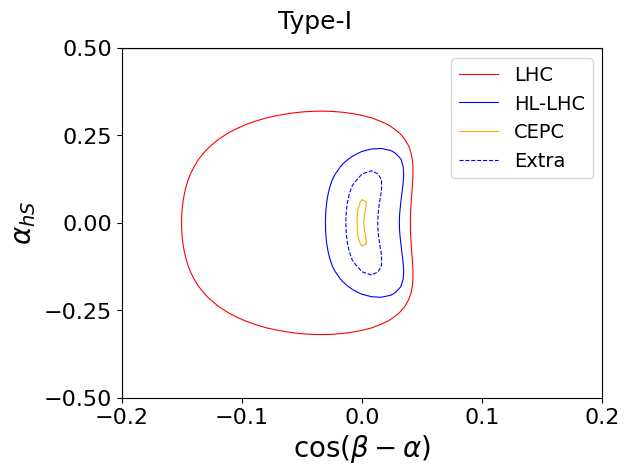}
\includegraphics[width=0.49 \linewidth]{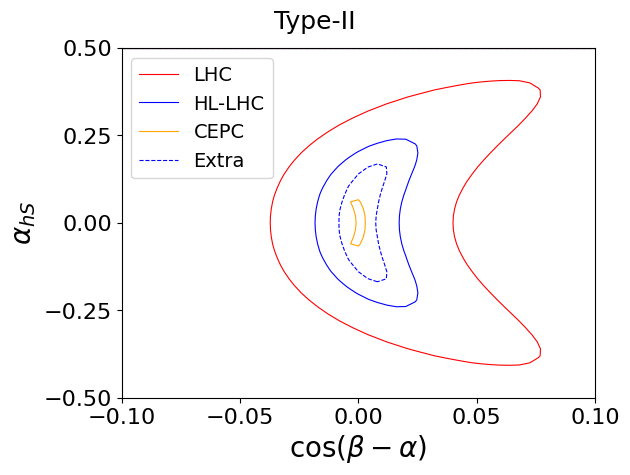}
\includegraphics[width=0.49 \linewidth]{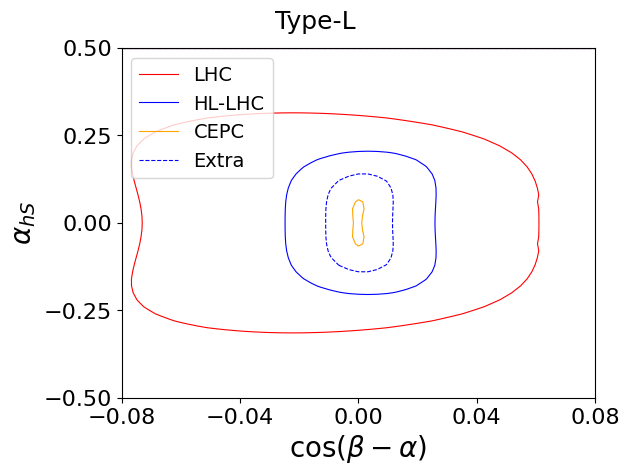}
\includegraphics[width=0.49 \linewidth]{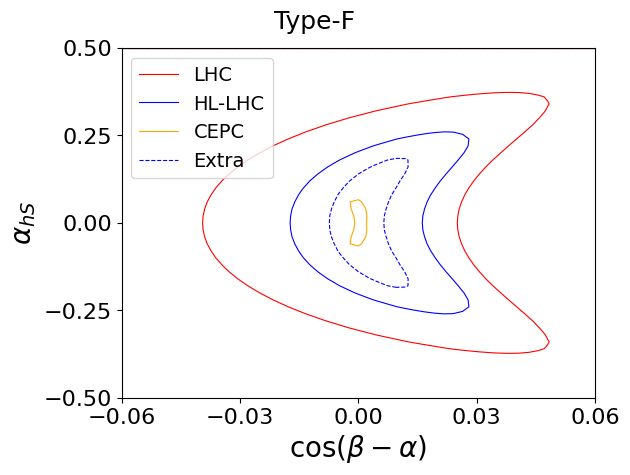}
\caption{The allowed regions at 95\% C.L. for the four types of 2HDMS in the plane of $\cos(\beta-\alpha)$ vs $\alpha_{hS}$. The red solid line is for the LHC precision measurement at 13 TeV with around 140 $fb^{-1}$. The blue solid line is for the HL-LHC precision measurement at 14 TeV with 3000 $fb^{-1}$ while the blue dashed line is for the extrapolated HL-LHC. The yellow solid line is for the CEPC precision measurement at a center-of-mass energy of 240 GeV with 20 $ab^{-1}$.}
    \label{fig:cbaa1}
\end{figure}

Now we want to consider the contribution from the singlet Higgs, shown as \autoref{fig:cbaa1} with $\tanb=1$. It is evident that the results for LHC, HL-LHC, and CEPC all form closed regions.

For the Type-II results of 2HDMS, shown as the upper right panel of \autoref{fig:cbaa1}. We find that the SM couplings in \autoref{eq:coupling2} are symmetric around $\alpha_{hS}=0$ since it shows only with $\cos (\alpha_{hS})$. Therefore, for the LHC, HL-LHC, and CEPC, the allowed parameter space in Type-II are all symmetric around $\alpha_{hS}=0$. Similar behavior also appear in Type-I, Type-L and Type-F. For the Type-II, we also find that larger $\cos(\beta-\alpha)$ will allow larger $|\alpha_{hS}|$. It is because, for $c_{huu}$, when $\cos(\beta-\alpha)$ increases to the positive direction, $\lvert \alpha_{hS} \rvert$ must correspondingly increases to ensure the existence of the allowed region at 95\% C.L. based on \autoref{eq:coupling2}. For $c_{hdd}$ and $c_{hll}$, as $\cos(\beta-\alpha)$ increases to the negative direction, $\lvert \alpha_{hS} \rvert$ must correspondingly increase. The interactions of these couplings ultimately result in that for LHC result and HL-LHC result, $\lvert \alpha_{hS} \rvert$ will increase as $\cos(\beta-\alpha)$ increases to the positive direction. For CEPC result, $\lvert \alpha_{hS} \rvert$ will increase as $\cos(\beta-\alpha)$ increases to the negative direction. For HL-LHC, $\tau\tau$ coupling provides the strongest constraint to the range of $\cos(\beta-\alpha)$. $WW$ coupling provides the strongest constraint to the range of $\lvert \alpha_{hS} \rvert$. For CEPC, $gg$ coupling, $\tau\tau$ coupling and $bb$ coupling provide the strongest constraint to the range of $\cos(\beta-\alpha)$ and $ZZ$ coupling provides the strongest constraint to the range of $\lvert \alpha_{hS} \rvert$. For LHC, the maximum range of $\cos(\beta-\alpha)$ is (-0.038, 0.078), which for HL-LHC(-0.018, 0.026), for extrapolated HL-LHC(-0.008, 0.012), for CEPC(-0.003, 0.003). The maximum of $\lvert \alpha_{hS} \lvert$ is 0.410 for LHC, 0.242 for HL-LHC, 0.170 for extrapolated HL-LHC and 0.066 for CEPC.

For the Type-I results of 2HDMS, shown as the upper left panel of \autoref{fig:cbaa1}. Unlike the results in Type-II, in Type-I, due to the change in the down quark Yukawa couplings, the results for LHC, HL-LHC, and CEPC indicate that $\alpha_{hS}$ and $\cos(\beta-\alpha)$ are relatively free within the region, with no significant correlation observed. For HL-LHC, $\tau\tau$ coupling provides the strongest constraint to the range of $\cos(\beta-\alpha)$ based on \autoref{fig:kappahl} at the Appendix. $WW$ coupling provides the strongest constraint to the range of $\lvert \alpha_{hS} \rvert$. For CEPC, $gg$ coupling, $\tau\tau$ coupling and $bb$ coupling provide the strongest constraint to the range of $\cos(\beta-\alpha)$ and $ZZ$ coupling provides the strongest constraint to the range of $\lvert \alpha_{hS} \rvert$ based on \autoref{fig:kappacepc} at the Appendix. Here, for LHC, the maximum range of $\cos(\beta-\alpha)$ is (-0.152, 0.042), which for HL-LHC(-0.032, 0.034), for extrapolated HL-LHC(-0.015, 0.015), for CEPC(-0.006, 0.003). The maximum of $\lvert \alpha_{hS} \rvert$ is 0.321 for LHC, 0.213 for HL-LHC, 0.148 for extrapolated HL-LHC and 0.007 for CEPC.



Similar behavior happen in the result of Type-L and Type-F. The change of lepton Yukawa couplings caused a small difference.
To better clarify the constraints on the 2HDMS parameter space in $\alpha_{hS}$ vs $\cos(\beta-\alpha)$, a summary of the maximum and minimum permissible values of $\cos(\beta-\alpha)$ and $\alpha_{hS}$ is given in \autoref{tab:cbaa1}. Similarly, based on the solid blue line and the dashed blue line of~\autoref{fig:cbaa1}, it is clear that the upgraded analysis method for LHC is tremendous. The dashed blue line for extrapolated HL-LHC based on latest LHC works much better than HL-LHC and approaches the result of CEPC as the~\autoref{tab:cbaa1}. The maximum range of $\cos(\beta-\alpha)$ for HL-LHC is basically larger than twice that of extrapolated HL-LHC and the range of $\lvert \alpha_{hS} \rvert$ for HL-LHC is about 1.5 times larger than that of extrapolated HL-LHC. 

\begin{table}[h!] 
    \centering
    \begin{tabular}{*{2}{r|cccc}}
        \toprule 
         Type &LHC&HL-LHC&Extra&CEPC   \\   
        \midrule
        Type-I  $\cos(\beta-\alpha)$&(-0.152, 0.042)&(-0.032, 0.034)&(-0.015, 0.015)&(-0.006, 0.003)\\
        $\lvert \alpha_{hS} \rvert$ &0.321&0.213&0.148&0.007\\
        \hline
          Type-II $\cos(\beta-\alpha)$&(-0.038, 0.078)&(-0.018, 0.026)&(-0.008, 0.012)&(-0.003, 0.003)\\
          $\lvert \alpha_{hS} \rvert$ & 0.410&0.242&0.170&0.066\\
          \hline
          Type-L $\cos(\beta-\alpha)$&(-0.077, 0.062)&(-0.025, 0.026)&(-0.011, 0.012)&(-0.002, 0.002)\\
          $\lvert \alpha_{hS} \rvert$ &0.315&0.206&0.141&0.065\\
          \hline
           Type-F $\cos(\beta-\alpha)$&(-0.039, 0.049)&(-0.017, 0.028)&(-0.007, 0.013)&(-0.002, 0.003)\\
           $\lvert \alpha_{hS} \rvert$ & 0.372&0.261&0.186&0.067\\
          \midrule
    \end{tabular}
     
     \caption{Maximum allowed range for $\cos(\beta-\alpha)$ and $\lvert \alpha_{hS} \rvert$ at 95\% C.L..}
     \label{tab:cbaa1}
\end{table}
\subsection{$\tan\beta$ - $\alpha_{hS}$}
\begin{figure}[h]
    \centering
    \includegraphics[width=0.5\linewidth]{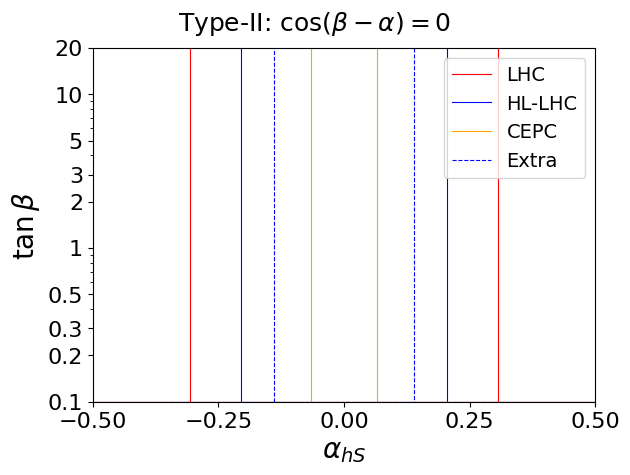}
    \caption{The allowed regions at 95\% C.L. for Type-II in the plane of $\tan\beta$ vs $\alpha_{hS}$ with $\cos(\beta-\alpha)=0$. The red line is for LHC measurement, while the blue line is for HL-LHC and the yellow line is for CEPC.}
    \label{fig:type2_tba1}
\end{figure}

Finally we want to study the genearal relationship of $\tan\beta$ - $\alpha_{hS}$ in 2HDMS parameter space. For the alignment limit $\cos(\beta-\alpha)=0$ scenario, shown as \autoref{fig:type2_tba1}, we present the results for Type-II in 2HDMS. The maximal range of $\alpha_{hS}$ is 0.305 for LHC, 0.203 for HL-LHC, 0.139 for extrapolated HL-LHC and 0.066 for CEPC. At this stage, \autoref{eq:coupling3} shows that SM couplings are only related to $\alpha_{hS}$. It means that regardless of the value of $\tan\beta$, it has virtually no impact on the parameter space of 2HDMS. Similar results are observed across other types of 2HDMS models, we will not elaborate further here. 

\begin{figure}[htpb]
  \centering
\includegraphics[width=0.49 \linewidth]{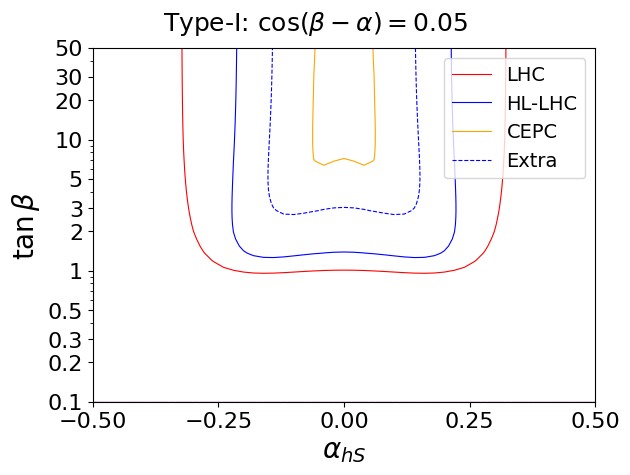}
\includegraphics[width=0.49 \linewidth]{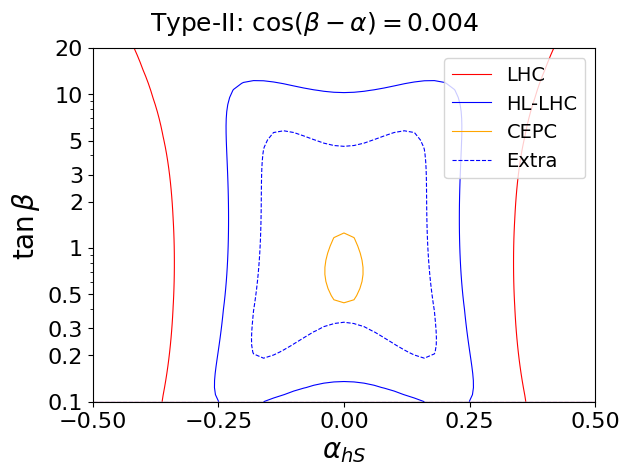}
\caption{The allowed regions at 95\% C.L. for the 2HDMS in the plane of $\tanb$ vs $\alpha_{hS}$. The red solid line is for LHC. The blue solid line is for HL-LHC while the blue dashed line is for the extrapolated HL-LHC. The yellow solid line is for the CEPC. The left panel is for Type-I with $\cos(\beta-\alpha) = 0.05$ while the right panel is for Type-II with $\cos(\beta-\alpha) = 0.004$.}
    \label{fig:tba1}
\end{figure}

In addition, we also study the scenario deviated from the $\cos(\beta-\alpha) \neq 0$ to study the relation between $\tan\beta$ and $\alpha_{hS}$. \autoref{fig:tba1} shows the result of 2HDMS in Type-I and Type-II. For the Type-I result, based on the upper left panel of \autoref{fig:cbatb}, the allowed parameter space at small $\tanb$ is strongly constrained. As a result, when $\cos(\beta-\alpha) \neq 0$, the allowed parameter space exists only in $\tanb > 1$. All the couplings are $\cot\beta$-enhanced, as shown by the upper left panel of \autoref{fig:cbatb} and \autoref{eq:coupling3} at large $\tanb$, thus the allowed parameter space is basically unconstrained even when $\cos(\beta-\alpha) \neq 0$. Therefore, for the LHC, HL-LHC and CEPC results, at large $\tanb$, no closed range is formed. Similarly, at large $\tanb$, the value of $\tanb$ has virtually no import on $\alpha_{hS}$. For the Type-II result, the allowed parameter space forms a closed region. However, the value of $\tanb$ has no significant effect on $\alpha_{hS}$. No further research will be conducted on $\tanb$ vs $\alpha_{hS}$.

Based on \autoref{fig:type2_tba1} and \autoref{fig:tba1}, the dashed blue line for extrapolated HL-LHC has the same properties as HL-LHC and works much better than HL-LHC.

\section{Light Higgs}
\label{sec5}
%
\subsection{Light Higgs contribution}
\begin{figure}[htpb]
  \centering
\includegraphics[width=0.49 \linewidth]{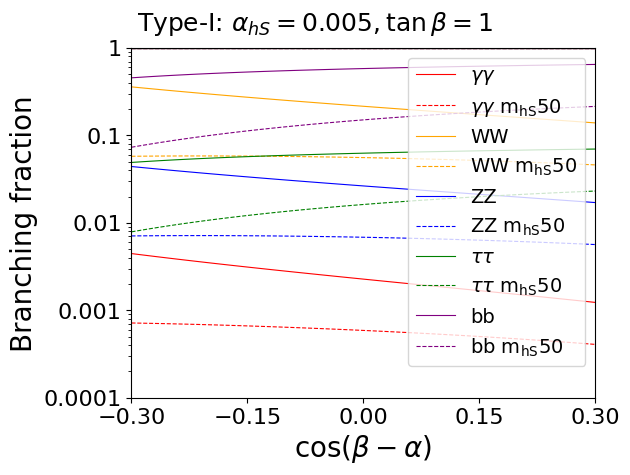}
\includegraphics[width=0.49 \linewidth]{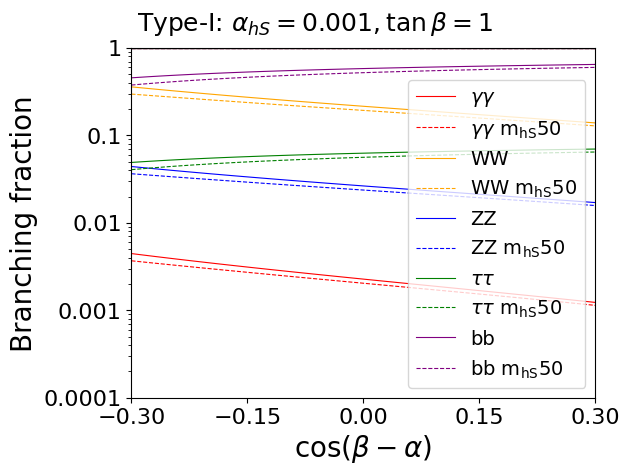}
\includegraphics[width=0.49 \linewidth]{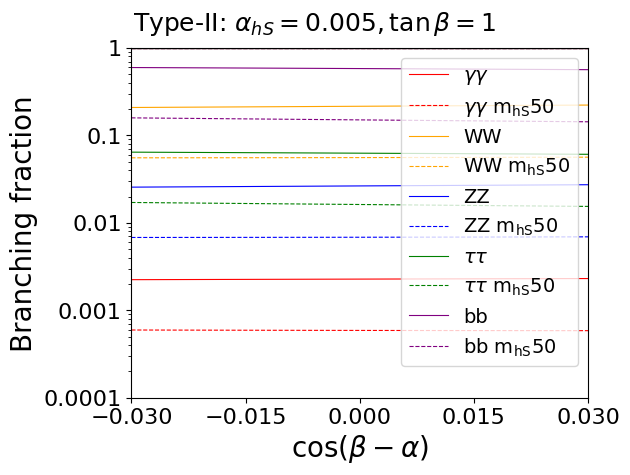}
\includegraphics[width=0.49 \linewidth]{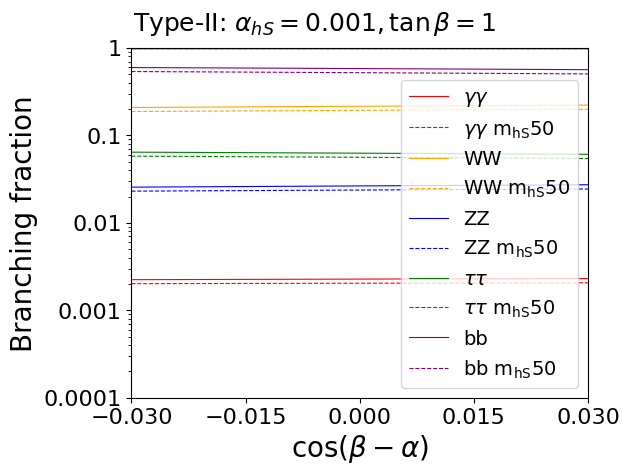}
\caption{The branching fraction of different decay channels for 2HDMS with $\alpha_{hS}=0.005$ or $\alpha_{hS}=0.001$. The two upper panels are for the result of Type-I and the two lower panels are for the result of Type-II. The solid line is for 2HDMS with  $m_{hS}$=125GeV, while the dashed line is for 2HDMS with light Higgs($m_{hS}$=50GeV).}
    \label{fig:br}
\end{figure}

In Higgs physics, a light Higgs typically refers to an additional neutral scalar or pseudoscalar particle whose mass is significantly below that of the 125 GeV Higgs boson, with masses ranging from several tens of GeV down to well below 125 GeV. Light Higgs bosons arise naturally in BSM scenarios and may be probed experimentally. In the 2HDMS model, when a light Higgs exists, the 125 GeV SM-like Higgs boson can decay into two light Higgs bosons. It accounts for a large fraction of the Higgs decay width. This indirectly leads to a reduction in the other decay channels such as $bb$, $\gamma\gamma$. In $\alpha_{hS}$ vs $\cos(\beta-\alpha)$ with $\tanb=1$, the branching fraction for the 125 GeV SM-like Higgs boson decay into two light Higgs is as high as 99\% if $\lvert \alpha_{hS} \rvert > 0.04$. It is hard for us to study the other decay channels and parameter space of 2HDMS. \autoref{fig:br} shows the effect of varying the value of $\alpha_{hS}$ on reducing the impact of the light Higgs decay channel for 2HDMS in the Type-I or Type-II with small $\alpha_{hS}$. Comparing the left and right panels of \autoref{fig:br}, it is clear that as the value of $\alpha_{hS}$ decreases, the effect of the light Higgs decay channel on other decay channels becomes smaller. We also find that when $\alpha_{hS}=0.001$, the effect of the light Higgs on other decay channels is extremely small. To better study the impact of the light Higgs on 2HDMS, $\text{Br}( h\to h_S h_S)$ should be lower than 10\%. Therefore, to ensure that the light Higgs has no significant impact on the 2HDMS under current constraints, the value of $\alpha_{hS}$ is restricted to below 0.001 in all four types with the range of $\cos(\beta - \alpha)$ and $\tanb$ essentially unchanged. 
%
\subsection{light Higgs results}
\subsubsection{$\alpha_{hS}$ - $\cos(\beta - \alpha)$ in light Higgs}
\begin{figure}[htpb]
  \centering
\includegraphics[width=0.49 \linewidth]{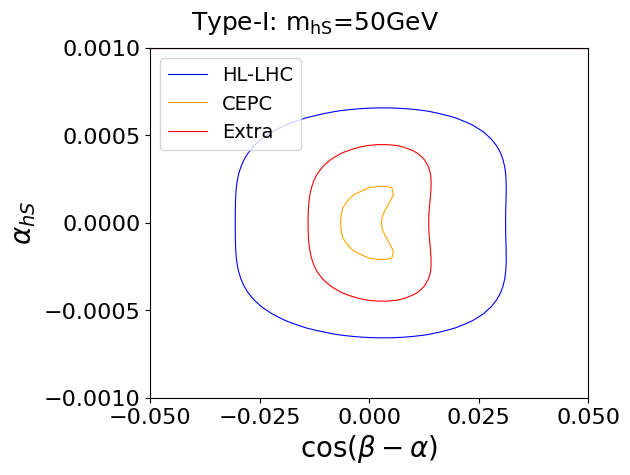}
\includegraphics[width=0.49 \linewidth]{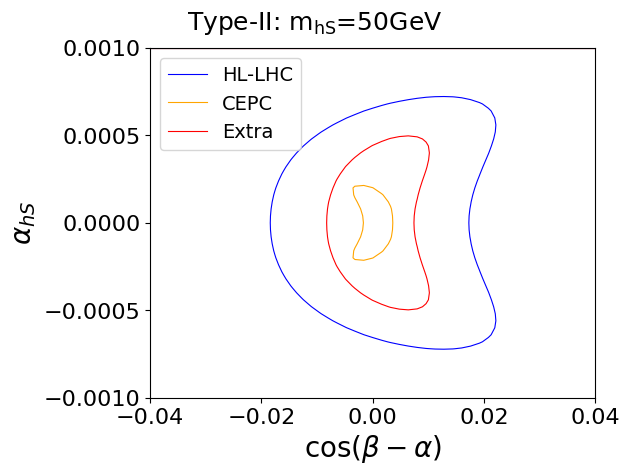}
\includegraphics[width=0.49 \linewidth]{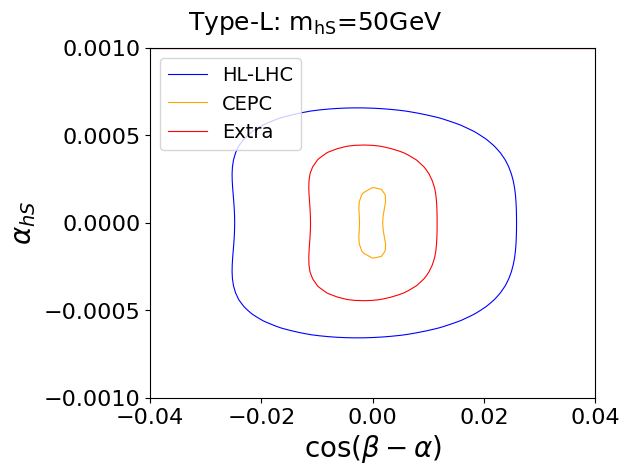}
\includegraphics[width=0.49 \linewidth]{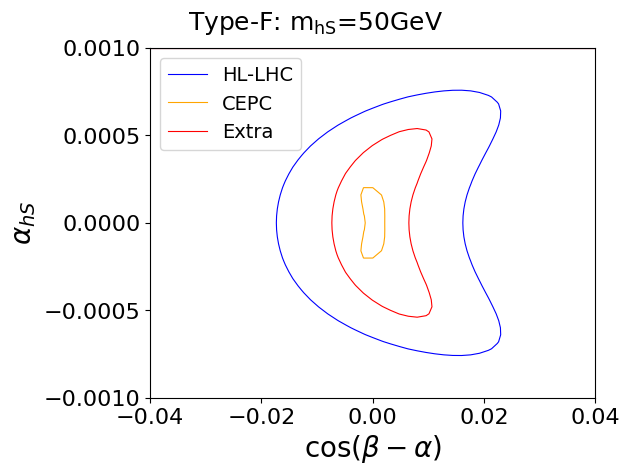}
\caption{The allowed regions at 95\% C.L. for the four types of 2HDMS with light Higgs($m_{hS}$=50GeV) in the plane of $\cos(\beta-\alpha)$ vs $\alpha_{hS}$. The blue line is for HL-LHC precision measurement at 14 TeV with 3000 $fb^{-1}$ and the red line is for extrapolated HL-LHC precision measurement. The yellow line is for the CEPC precision measurement at a center-of-mass energy of 240 GeV with 20 $ab^{-1}$.}
    \label{fig:l50_cbaa1}
\end{figure}

\autoref{fig:l50_cbaa1} shown the 95\%C.L. allowed regions in $\alpha_{hS}$ vs $\cos(\beta-\alpha)$ with $\tanb=1$ for various types of 2HDMS with light Higgs($m_{hS}$=50GeV). Light Higgs in 2HDMS provides strongly constraint to 2HDMS parameter space.

In Type-I, shown as the upper left panel of \autoref{fig:l50_cbaa1}. It is similar to 2HDMS without light Higgs in Type-I that for the HL-LHC, HL-LHC extrapolation and CEPC results, $\alpha_{hS}$ and $\cos(\beta-\alpha)$ are relatively free within the region, with no significant correlation observed. The allowed parameter space in Type-I are all symmetric around $\alpha_{hS}=0$. For HL-LHC, $\tau\tau$ coupling provides the strongest constraint to the range of $\cos(\beta-\alpha)$. $WW$ coupling provides the strongest constraint to the range of $\alpha_{hS}$. For CEPC, $\tau\tau$ channel provides the strongest constraint to the range of $\cos(\beta-\alpha)$ and $ZZ$ coupling provides the strongest constraint to the range of $\alpha_{hS}$. For HL-LHC, the maximum range of $\cos(\beta-\alpha)$ is (-0.0307, 0.0318), which for extrapolated HL-LHC(-0.0140, 0.0144) and for CEPC(-0.0066, 0.0056). The maximum of $\lvert \alpha_{hS} \rvert$ times $10^4$ is 6.55 for HL-LHC, 4.49 for extrapolated HL-LHC and 2.07 for CEPC.

In Type-II, shown as the upper right panel of \autoref{fig:l50_cbaa1}. Similar to the 2HDMS without light Higgs in Type-II, the allowed parameter space are all symmetric around $\alpha_{hS}=0$. For HL-LHC result, $\lvert \alpha_{hS} \rvert$ will increase as $\cos(\beta-\alpha)$ increases to the positive direction. For CEPC result, $\lvert \alpha_{hS} \rvert$ will increase as $\cos(\beta-\alpha)$ increases to the negative direction. For HL-LHC, $\tau\tau$ coupling provides the strongest constraint to the range of $\cos(\beta-\alpha)$. $WW$ coupling provides the strongest constraint to the range of $\alpha_{hS}$. For CEPC, $gg$ coupling, $\tau\tau$ coupling and $bb$ coupling provide the strongest constraint to the range of $\cos(\beta-\alpha)$ and $ZZ$ coupling provides the strongest constraint to the range of $\alpha_{hS}$. For HL-LHC, the maximum range of $\cos(\beta-\alpha)$ is (-0.019, 0.022), which for extrapolated HL-LHC(-0.008, 0.011) and for CEPC(-0.003, 0.003). The maximum range of $\lvert \alpha_{hS} \rvert$ times $10^4$ is 7.22 for HL-LHC, 4.96 for extrapolated HL-LHC and 2.11 for CEPC.  




For the Type-L and Type-F models, they show the similar features as the models with $m_H > 125.5$ GeV, except for shrunk $\alpha_{hS}$ as Type-I and Type-II models. Finally here to better analyze the constraints imposed on the 2HDMS parameter space, we present the maximum and minimum permissible values of $\cos(\beta-\alpha)$ and $\alpha_{hS}$ for both non-light Higgs and light Higgs scenarios in the 2HDMS model, shown as \autoref{tab:cba_low}.

\begin{table}[h!] 
    \centering
    \begin{tabular}{*{2}{r|ccc}}
        \toprule 
         Type &HL-LHC&Extra&CEPC   \\   
        \midrule
        Type-I  $\cos(\beta-\alpha)$&(-0.0307, 0.0318)&(-0.0140, 0.0144)&(-0.0066, 0.0056)\\
        $\lvert \alpha_{hS} \rvert (\times e^{-4})$ &6.55&4.49&2.07\\
        \hline
          Type-II $\cos(\beta-\alpha)$&(-0.019, 0.022)&(-0.008, 0.011)&(-0.003, 0.003)\\
          $\lvert \alpha_{hS} \rvert (\times e^{-4})$ &7.22&4.96&2.11 \\
          \hline
          Type-L $\cos(\beta-\alpha)$&(-0.025, 0.025)&(-0.012, 0.011)&(-0.003, 0.002) \\
          $\lvert \alpha_{hS} \rvert (\times e^{-4})$ &6.55&4.46&1.99\\
          \hline
           Type-F $\cos(\beta-\alpha)$&(-0.017, 0.023)&(-0.007, 0.011)&(-0.002, 0.002)\\
           $\lvert \alpha_{hS} \rvert (\times e^{-4})$ &7.61&5.42&2.01 \\
          \midrule
    \end{tabular}
     \caption{Maximum allowed range for $\cos(\beta-\alpha)$ and $\alpha_{hS}$ at 95\% C.L. in light Higgs.}
     \label{tab:cba_low}
\end{table}

\subsubsection{$\tan\beta$ - $\cos(\beta-\alpha)$ in light Higgs}
\begin{figure}[htpb]
  \centering
\includegraphics[width=0.49 \linewidth]{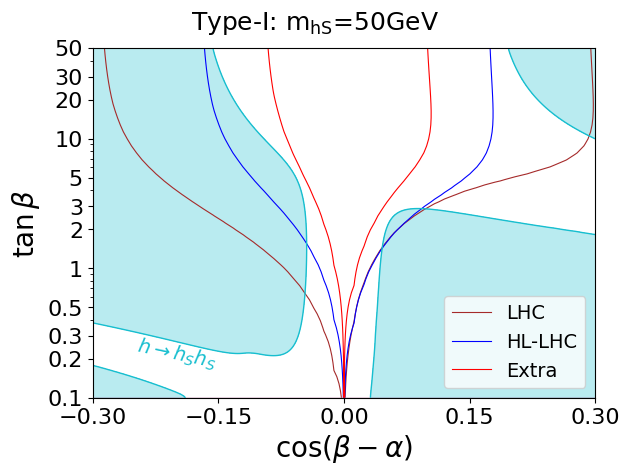}
\includegraphics[width=0.49 \linewidth]{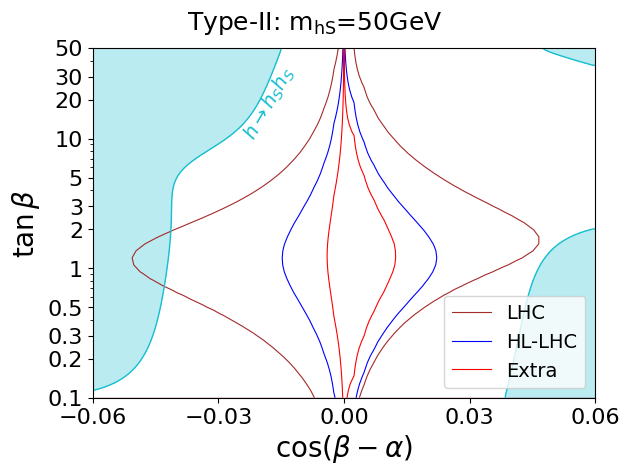}
\includegraphics[width=0.49 \linewidth]{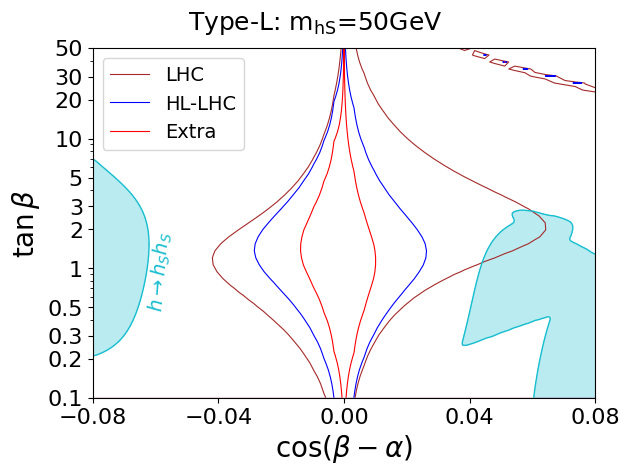}
\includegraphics[width=0.49 \linewidth]{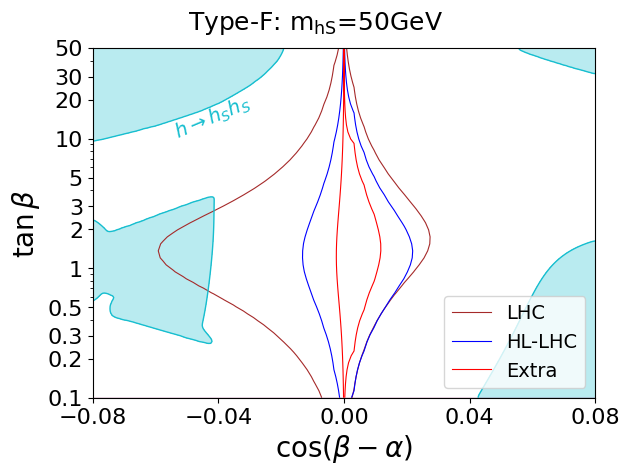}
\caption{The result from direct searches and the allowed regions at 95\% C.L. for the four types of 2HDMS with light Higgs($m_{hS}$=50GeV) in the plane of $\cos(\beta-\alpha)$ vs $\tan\beta$($\alpha_{hS}=0.0004$).  The brown line is for the result of LHC precision measurement at 13 TeV with 140 $fb^{-1}$. The blue line is for HL-LHC precision measurement at 14 TeV with 3000 $fb^{-1}$ while the red line is for extrapolated HL-LHC. The cyan regions are excluded at 95\% C.L. by the direct search $h \to h_S h_S$.}
\label{fig:l50_cbatb}
\end{figure}

As shown in the plane that $\alpha_{hS} $ vs $ \cos(\beta-\alpha)$, when a light Higgs boson exists in 2HDMS, the parameter space for CEPC becomes severely constrained, with the absolute value of $\alpha_{hS}$ limited to less than 0.00025. Here we set the value of $\alpha_{hS}=0.0004$ and focus on the restrictions of LHC and HL-LHC parameter space. To involve the light $h_S$, the results of LHC are produced by our own program with the latest LHC precisions \cite{ATLAS:2026pdi,CMS:2026zxu}. In this case, the new direct searches for the 125 GeV SM-like Higgs boson decaying into two light singlet Higgs bosons($m_{hS}$=50GeV) comes out.
\autoref{fig:l50_cbatb} shown the result from direct searches and the 95\%C.L. allowed regions in $\tanb$ vs $\cos(\beta-\alpha)$ for various types of 2HDMS with light Higgs($m_{hS}$=50GeV).

In Type-II, shown as the upper right panel of \autoref{fig:l50_cbatb}. At large $\tanb$ and small $\tanb$, 2HDMS with light Higgs are highly similar to the 2HDMS that the allowed parameter space reduce significantly. For HL-LHC with light Higgs, at large $\tanb$, $\tau\tau$ coupling provides the strongest constraint. At small $\tanb$, $\gamma\gamma$ coupling rules out large $\cos(\beta-\alpha)$. Due to the reduced allowed parameter space at large $\tanb$ and small $\tanb$, the interval gets is maximal region around $\tanb=1$, which for LHC(-0.050, 0.047), HL-LHC(-0.015, 0.022) and extrapolated HL-LHC(-0.004, 0.0012). At this scenario, the allowed parameter space for 2HDMS with light Higgs does not show a significant reduction compared to that of 2HDMS without light Higgs. However, based on the upper right panel of \autoref{fig:l50_cbaa1}, for 2HDMS with light Higgs in Type-II, $\lvert \alpha_{hS} \rvert$ will increase as $\cos(\beta-\alpha)$ increases to the positive direction. Thus when $\alpha_{hS}=0.0004$, the value of $\cos(\beta-\alpha)$ will increase to the positive direction, and its allowed parameter space will correspondingly shift to the positive direction of the $\cos(\beta-\alpha)$. Comparing the result of direct searches and global fit, we can see that at both large $\tanb$ and small $\tanb$, the global fit works better than the direct searches. Similar behavior also happen in Type-L and Type-F.

In Type-I, shown as the upper left panel of \autoref{fig:l50_cbatb}. For LHC in light Higgs, $\lvert \cos(\beta-\alpha) \rvert$ is constrained to be less than about 0.29. For HL-LHC in light Higgs, $\lvert \cos(\beta-\alpha) \rvert$ is further constrained to be less than 0.18 while for extrapolated HL-LHC, $\lvert \cos(\beta-\alpha) \rvert$ is constrained to be less than 0.11. It is similar to 2HDMS without light Higgs in Type-I that the parameter space of 2HDMS with light Higgs will face stricter constraints at small $\tanb$. For HL-LHC, at small $\tanb$, $\tau\tau$ coupling provides the strongest constraint. The interval gets is maximal region around $\tanb>10$, which for LHC(-0.28, 0.29), HL-LHC(-0.16, 0.18) and extrapolated HL-LHC(-0.09, 0.11). For LHC, HL-LHC and extrapolated HL-LHC, their results in Type-I are all approximately symmetric around $\cos(\beta-\alpha)=0$ again. Compared to the case without $m_{h_S} <m_h/2$, the allowed parameter space for 2HDMS with light Higgs shows a significant reduction. 

The more important different comes from the SM-like Higgs exotic decay, $h \to h_S h_S$. As the cyan region shows, this direct search $h \to h_S h_S$ can cover large parts at large $\tanb$, where is weakly explored by the global fit . As a result, the direct searches show the complementary probe ability to the Higgs precision measurements.


In Type-L, shown as the lower left panel of \autoref{fig:l50_cbatb}. At large $\tanb$ and small $\tanb$, it is similar to that of Type-II that the allowed parameter space reduce significantly. At large $\tanb$, $\tau\tau$ coupling provides the strongest constraint. $\gamma\gamma$ coupling provides the strongest constraint at small $\tanb$. In Type-L, the allowed parameter space for 2HDMS with light Higgs does not show a significant reduction compared to that of 2HDMS without light Higgs. Similar to Type-I, the allowed parameter space are approximately symmetric around $\cos(\beta-\alpha)=0$.

In Type-F, shown as the lower right panel of \autoref{fig:l50_cbatb}. At large $\tanb$ and small $\tanb$, 2HDMS with light Higgs are highly similar to the 2HDMS that the allowed parameter space reduce significantly. For HL-LHC, at large $\tanb$, $bb$ coupling provides the strongest constraint and $\tau\tau$ coupling provides the strongest constraint at small $\tanb$. Similar to Type-II that the allowed parameter space for 2HDMS with light Higgs does not show a significant reduction compared to that of 2HDMS without light Higgs and its region will shift in the positive direction of the $\cos(\beta-\alpha)$.

To better analyze the constraints imposed on the 2HDMS parameter space in light Higgs, a summary of the maximum and minimum permissible values of $\cos(\beta-\alpha)$ for light Higgs scenarios in 2HDMS is given in \autoref{tab:cba_range}.

\begin{table}[h!] 
    \centering
    \begin{tabular}{*{2}{c|ccc}}
        \toprule 
         Type &LHC&HL-LHC&Extra   \\   
        \midrule
         Type-I &(-0.28, 0.29)&(-0.16, 0.18)&(-0.09, 0.11) \\
          Type-II &(-0.050, 0.047)&(-0.015, 0.022)&(-0.004, 0.012)\\
          Type-L &(-0.042, 0.064)&(-0.029, 0.026)&(-0.014, 0.010) \\
           Type-F &(-0.058, 0.027)&(-0.013, 0.021)&(-0.003, 0.012)\\
          \midrule
    \end{tabular}
    \caption{Maximum allowed range for $\cos(\beta-\alpha)$ at 95\% C.L. in light Higgs. }
    \label{tab:cba_range}
\end{table}

\section{Collider Comparison}
\label{sec6}

For the future $e^+ e^-$ collider, CEPC, FCC-ee and ILC, with the center-of-mass energy of 240-250 GeV, the dominant channel to measure the Higgs boson properties is the Higgsstrahlung process of $e^+ e^-$ $\rightarrow$ hZ. Here, we use the accuracy of FCC-ee and ILC for the study and the results are as follows.
\begin{figure}[htpb]
  \centering
\includegraphics[width=0.49 \linewidth]{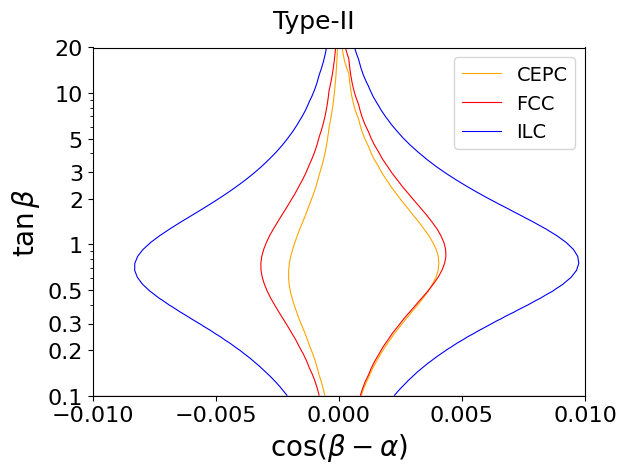}
\includegraphics[width=0.49 \linewidth]{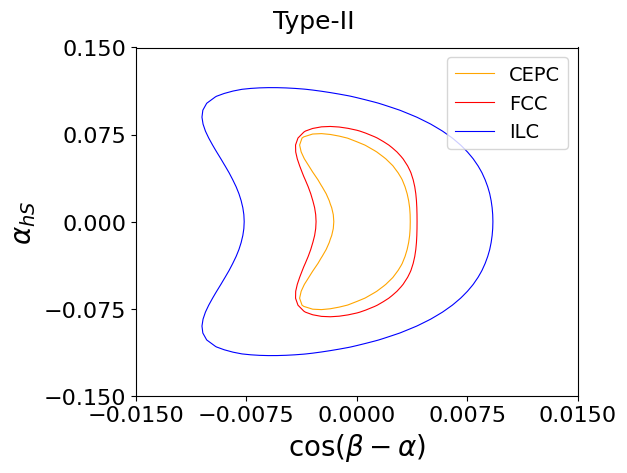}
\caption{The left panel is the allowed ranges at 95\% C.L. under different $e^+e^-$ colliders in the plane of $\cos(\beta-\alpha)$ vs $\tan\beta$. The right panel is the allowed ranges at 95\% C.L. under different $e^+e^-$ colliders in the plane of $\cos(\beta-\alpha)$ vs $\alpha_{hS}$. Here, the yellow line is for CEPC precision measurement at a center-of-mass energy of 240 GeV with 20 ab$^{-1}$. The red line is for FCC-ee precision measurement at a center-of-mass energy of 240 GeV with 10.8 ab$^{-1}$. The blue line is for ILC precision measurement at a center-of-mass energy of 250 GeV with 2 ab$^{-1}$.}
    \label{fig:ee_collider}
\end{figure}
These colliders accelerate particles to comparable high energies: 240 GeV for CEPC and FCC-ee, and 250 GeV for ILC. However, their integrated luminities differ significantly where CEPC at 20 ab$^{-1}$, FCC at 10.8 ab$^{-1}$, while ILC stands at merely 2 ab$^{-1}$. The results from both $\cos(\beta-\alpha)$ vs $\tan\beta$ and $\cos(\beta-\alpha)$ vs $\alpha_{hS}$ experiments demonstrate that although the particle energies at the $e^+e^-$ collider are similar, significant differences in beam luminosity among CEPC, FCC-ee and ILC lead to substantial variations in experimental precision. Among them, the experimental accuracy of CEPC and FCC-ee is significantly superior to that of ILC. Meanwhile, this disparity is further evident in the results at 95\% C.L.: those for CEPC and FCC-ee are smaller than that for ILC. Overall, whether considering $\cos(\beta-\alpha)$ vs $\tan\beta$ or $\cos(\beta-\alpha)$ vs $\alpha_{hS}$, The results of CEPC are better than those of the FCC, and the results of FCC-ee are better than those of the ILC. For FCC-ee and ILC, they exhibit the same trend of variation as CEPC (as mentioned earlier), particularly in both $\cos(\beta-\alpha)$ vs $\tan\beta$ and $\cos(\beta-\alpha)$ vs $\alpha_{hS}$. For FCC-ee and ILC in $\cos(\beta-\alpha)$ vs $\alpha_{hS}$, $gg$ coupling, $bb$ coupling and $\tau\tau$ coupling provide the strongest constraint to the range of $\cos(\beta-\alpha)$. $ZZ$ coupling provides the strongest constraint to the range of $\alpha_{hS}$. For FCC-ee and ILC in $\cos(\beta-\alpha)$ vs $\tan\beta$, at large $\tanb$, $bb$ coupling provides the strongest constraint and $gg$ coupling provides the strongest constraint at small $\tanb$. Noted that for FCC-ee and CEPC, their results are all approximately symmetric around $\alpha_{hS}=0$ in $\cos(\beta-\alpha)$ vs $\alpha_{hS}$ and approximately symmetric around $\cos(\beta-\alpha)=0$ in $\cos(\beta-\alpha)$ vs $\tan\beta$. We have observed that, for the $\cos(\beta-\alpha)$ vs $\tan\beta$ scenario, in the lower-right corner of both CEPC and FCC-ee regions, the FCC-ee results are slightly superior to those of CEPC. This is due to the significant impact of $gg$ and $cc$ channel accuracy on the lower-right corner regions of CEPC and FCC-ee. The accuracy of the $gg$ and $cc$ channels in CEPC is inferior to that of FCC-ee, ultimately leading to this outcome.

\section{Summary}

After the discovery of the SM-like Higgs boson at the LHC, exploring the Higgs sector becomes the frontier of particle physics. In this work, we studied the implications of the Higgs precision measurements from various Higgs factories (LHC, HL-LHC, CEPC, ILC, FCC-ee) on new physics models. In particular, we focused
on the singlet extension of Two Higgs Doublet Model, which can involve a richer spectrum of phenomenological signatures, such as additional DM candidate compared to the 2HDM. We performed the global fit with SM-like Higgs precisions for four types of 2HDMS. For the scenario with light singlet Higgs $m_{h_S} < m_h$, we also studied the specific direct searches $h \to h_S h_S$ at LHC, and found it could work better in the large $\tanb$ region than the fit results. 

First, after a general description about the 2HDMS at \autoref{sec2}, we listed the latest (expected) accuracies on  the Higgs factories at \autoref{sec3}. During our study, it is found that the latest LHC precisions may work better in some regions than HL-LHC data at 2019. We supposed the results come from better performance at updated LHC, thus we also had the extrapolated study for HL-LHC 3000 $fb^{-1}$ based on the latest LHC performance around 140 $fb^{-1}$. 
 
Our main results start from \autoref{sec4}.
We investigated the parameter space of the 2HDMS model, focusing primarily on the constraints imposed by the usual 2HDM parameters $\cos(\beta-\alpha)$, $\tan\beta$, and the singlet-doublet mixing angel $\alpha_{hS}$ at 95\% C.L. at tree level. 
Meanwhile, we also shown the extrapolated HL-LHC precision to analyze the parameter space of 2HDMS and compared it with those of existing HL-LHC accuracy results. 
The first result locates in the plane $\cos(\beta-\alpha)$ vs $\tanb$, without singlet doublet mixing $\alpha_{hS}=0$, which can be called the 2HDM limit. Generally for Type-II, -F, -L models, both small and large $\tanb$ regions are strongly constrained, while Type-I is strongly restricted at small $\tanb$. $\cos(\beta-\alpha)$ regions show at \autoref{tab:cbatb}, and typically the values at Type-II and Type-I under LHC are less than 0.043 and 0.32. Once we have $\alpha_{hS}\neq 0$, the allowed regions for Type-II and -F would be shifted to the positive $\cos(\beta-\alpha)$ a little, while the regions for Type-I and -L are shrunk a little. 

For the 2HDMS in $\alpha_{hS}$ vs $\cos(\beta-\alpha)$, we noted that for Type-II, -F, as $\cos(\beta-\alpha)$ increases to the positive direction, the corresponding value of $\lvert \alpha_{hS} \rvert$ must correspondingly increase for LHC and HL-LHC, while it is contrary for CEPC. For Type-I, -L, it is different that $\alpha_{hS}$ and $\cos(\beta-\alpha)$ are relatively free, without any significant correlation observed.
For the 2HDMS in $\tanb$ vs $\alpha_{hS}$, the variation of $\tanb$ has no appreciable effect on $\alpha_{hS}$. Thus the precision measurement at HL-LHC deserve new updates.
We also found that for the extrapolated HL-LHC, the results will yield significantly better outcomes than the original results. Especially the ranges of the allowed $|\cos(\beta-\alpha)|$ of Type-II, -L, -F under extrapolated HL-LHC is about half of themselves under HL-LHC 2019. 

In \autoref{sec5}, we investigated the scenario in which light singlet-like Higgs $h_S$ exist in 2HDMS, where a SM-like Higgs boson may decay into two light Higgs bosons, and similarly analyzed how this scenario constrains the parameter space of 2HDMS. The first important difference is $\lvert \alpha_{hS} \rvert$ is strongly shrunk from 0.035 under HL-LHC to be less than 0.001 . In $\tan\beta$ vs $\cos(\beta-\alpha)$, the allowed parameter space will reduce slightly for Type-I. For Type-II, -F, the allowed parameter space will shift in the positive direction of the $\cos(\beta-\alpha)$. Another important difference locates at $h \to h_S h_S$ searches. For Type-I we found it strongly reduce the $|\cos(\beta-\alpha)|$ at large $\tanb$ regions, where the usual direct searches and fit results contribute weakly. 

We also studied three different future $e^+ e^-$ colliders: the CEPC, ILC and FCC-ee under the center-of-mass energy of 240-250 GeV. The results of CEPC and FCC-ee are obviously better than that of ILC with higher integrated luminosities.

\section*{Acknowledgements}
The authors would like to thank Juxiang Li and Shufang Su for useful discussions. CL, HZ and WS are supported by the Natural Science Foundation of China (NSFC) under grant numbers 12305115.
\section{Appendix: Individual contribution to global fit}
%
\begin{figure}[htpb]
  \centering
\includegraphics[width=0.4 \linewidth]{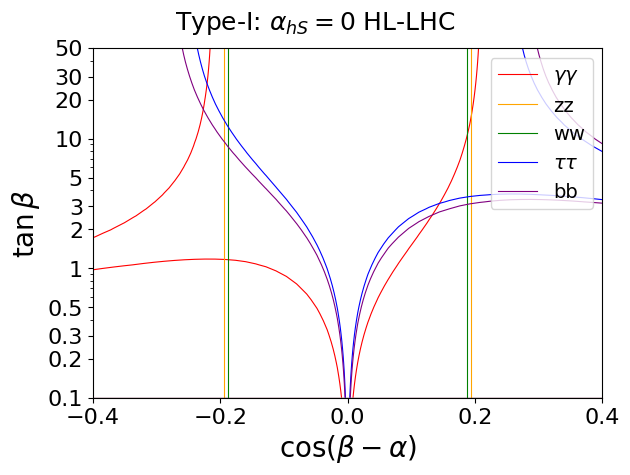}
\includegraphics[width=0.4 \linewidth]{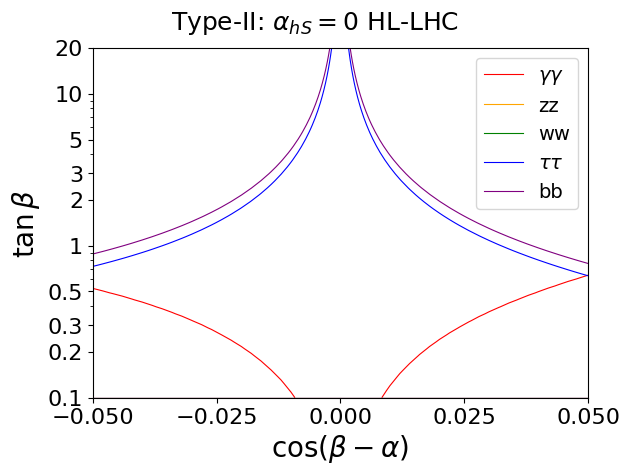}
\includegraphics[width=0.4 \linewidth]{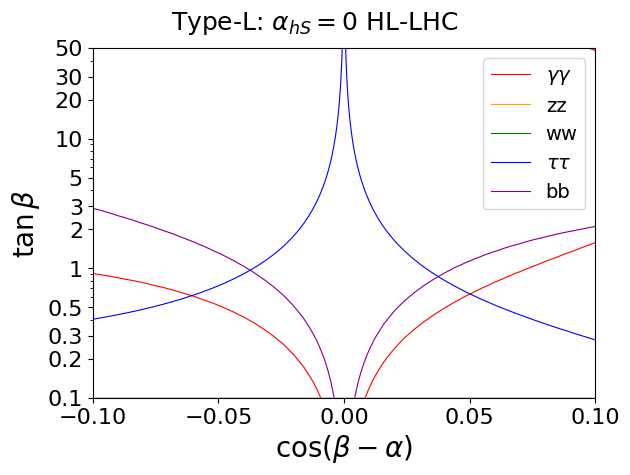}
\includegraphics[width=0.4 \linewidth]{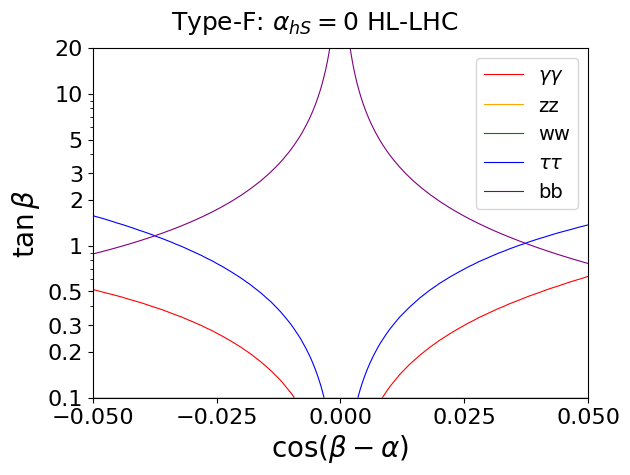}
\includegraphics[width=0.4 \linewidth]{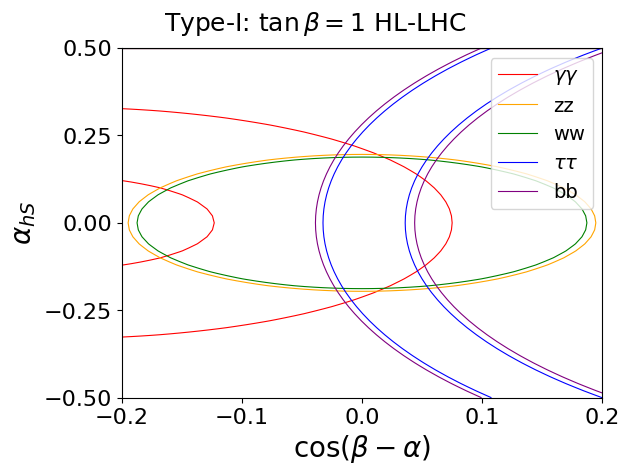}
\includegraphics[width=0.4 \linewidth]{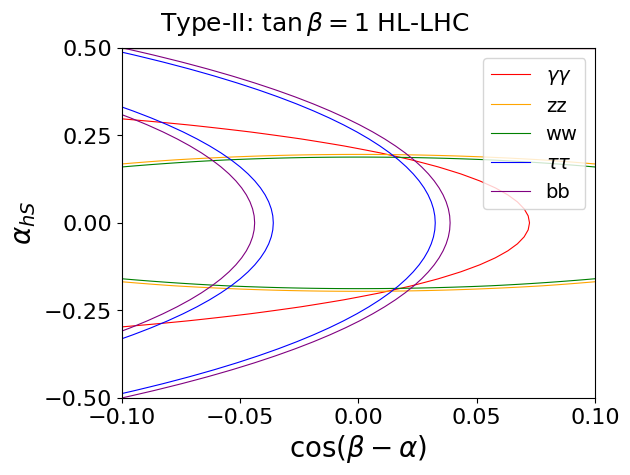}
\includegraphics[width=0.4 \linewidth]{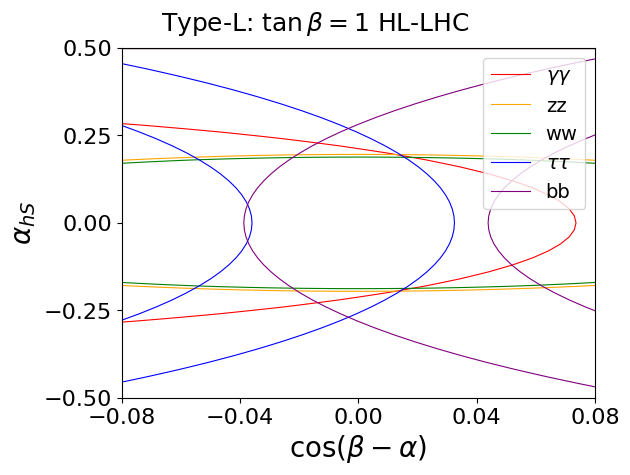}
\includegraphics[width=0.4 \linewidth]{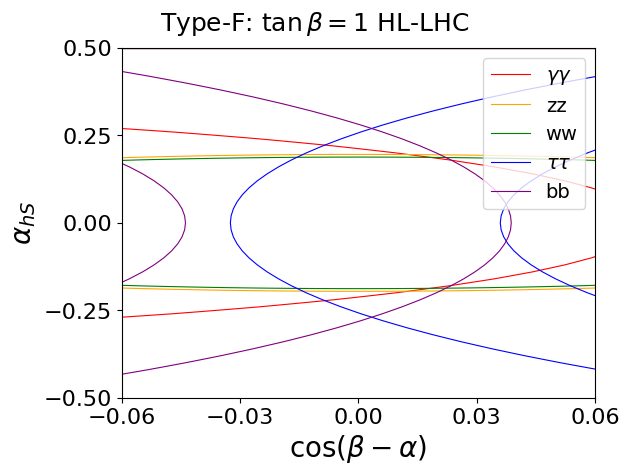}
\caption{The contributions of the respective kappa parameters to the global fit for HL-LHC.}
    \label{fig:kappahl}
\end{figure}
\begin{figure}[htpb]
  \centering
\includegraphics[width=0.4 \linewidth]{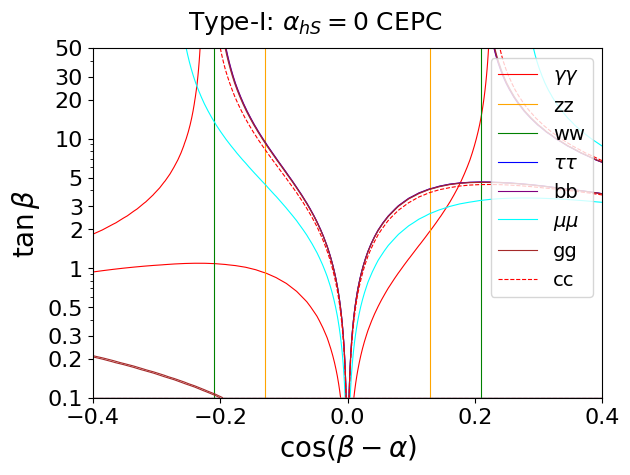}
\includegraphics[width=0.4 \linewidth]{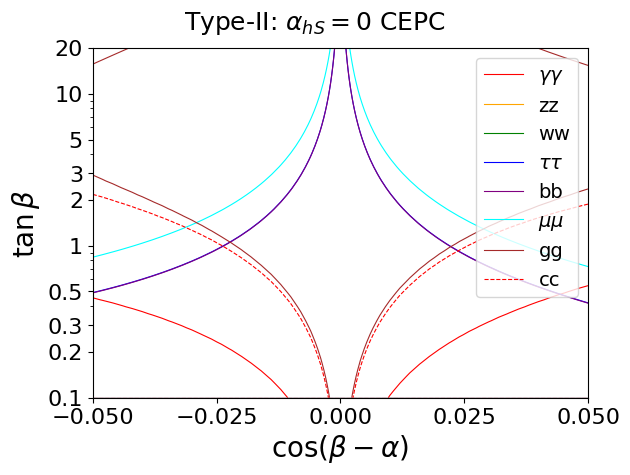}
\includegraphics[width=0.4 \linewidth]{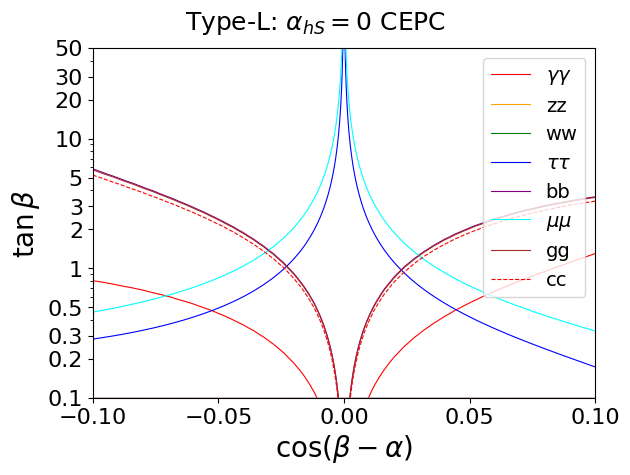}
\includegraphics[width=0.4 \linewidth]{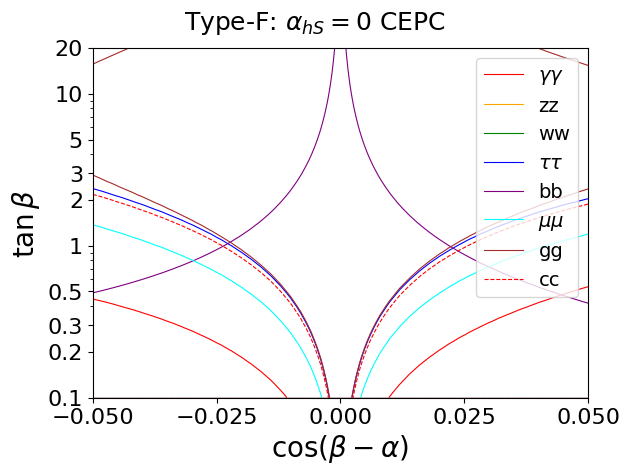}
\includegraphics[width=0.4 \linewidth]{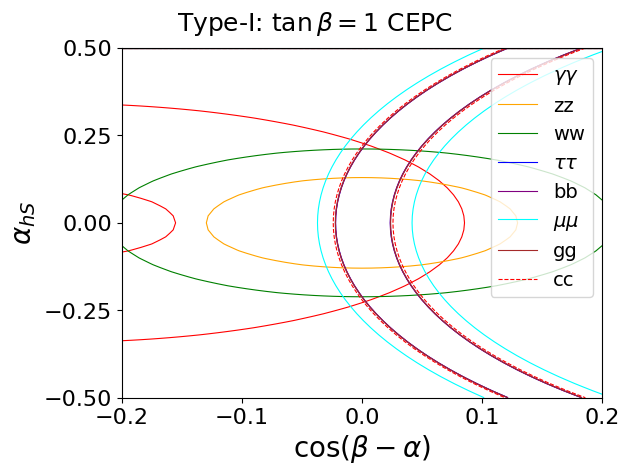}
\includegraphics[width=0.4 \linewidth]{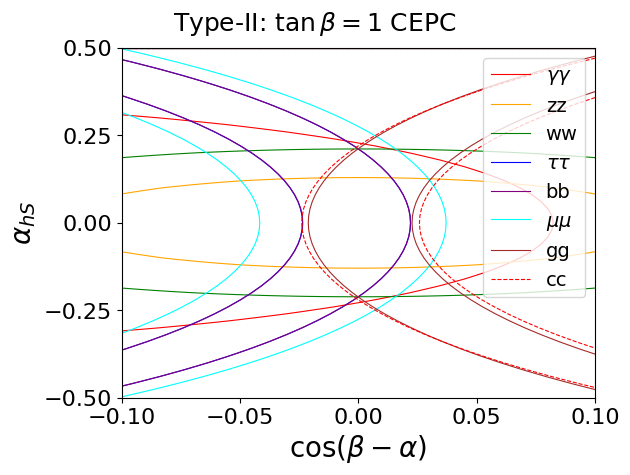}
\includegraphics[width=0.4 \linewidth]{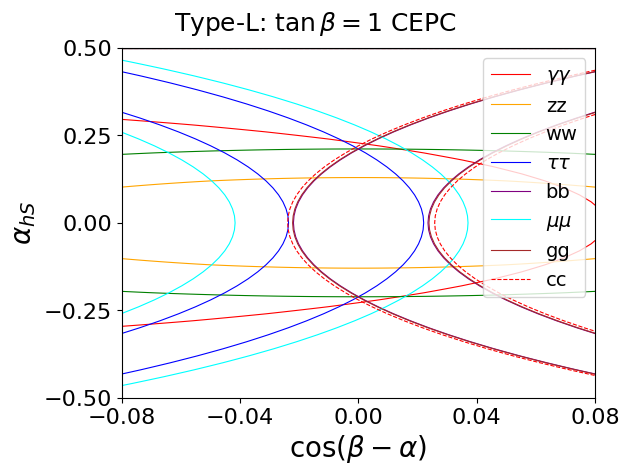}
\includegraphics[width=0.4 \linewidth]{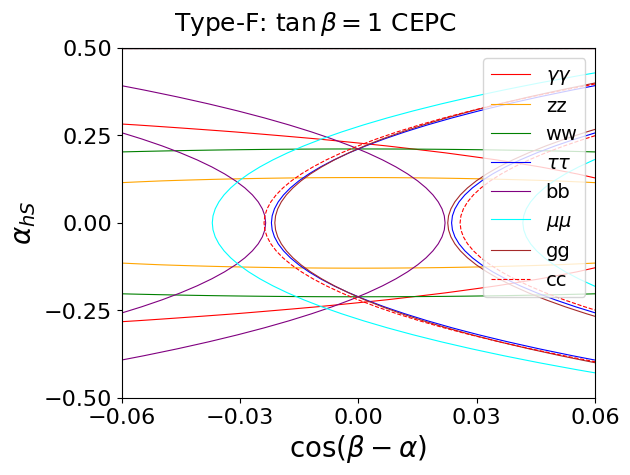}
\caption{The contributions of the respective kappa parameters to the global fit for CEPC.}
    \label{fig:kappacepc}
\end{figure}

\autoref{fig:kappahl} shows the channel constraints for HL-LHC in the $\cos(\beta-\alpha)$ vs $\tanb$ and $\cos(\beta-\alpha)$ vs $\alpha_{hS}$ plane, including $\gamma\gamma$(red line), $ZZ$(yellow line), $WW$(green line), $\tau\tau$(blue line) and $bb$(purple line). 
\autoref{fig:kappacepc} shows the channel constraints for CEPC in the $\cos(\beta-\alpha)$ vs $\tanb$ and $\cos(\beta-\alpha)$ vs $\alpha_{hS}$ plane, including $\gamma\gamma$(red line), $ZZ$(yellow line), $WW$(green line), $\tau\tau$(blue line), $bb$(purple line), $\mu\mu$(cyan line), $gg$(brown line) and $cc$(red dashed line). For \autoref{fig:kappahl} and \autoref{fig:kappacepc}, the four upper panels show the results of 2HDMS in 4 types in $\cos(\beta-\alpha)$ vs $\tanb$ and the four lower panels show the results of 2HDMS in 4 types in $\cos(\beta-\alpha)$ vs $\alpha_{hS}$.

\clearpage
\newpage
\bibliographystyle{JHEP}
\bibliography{ref_type1}

\end{document}